\documentclass[notitlepage,10pt,aps,prd,twocolumn,amsmath,amssymb,floatfix,superscriptaddress,nofootinbib,preprintnumbers]{revtex4-1}

\usepackage{amsfonts,amssymb,amsmath,graphicx,color,bm,epsfig}
\usepackage[normalem]{ulem}
\usepackage{multirow}

\definecolor{ultramarine}{rgb}{0.07, 0.04, 0.56}
\definecolor{cadmiumgreen}{rgb}{0.0, 0.42, 0.24}
\definecolor{indigo(dye)}{rgb}{0.0, 0.25, 0.42}
\usepackage[linktocpage=true]{hyperref}
\hypersetup{
colorlinks=true,
citecolor=ultramarine,
linkcolor=cadmiumgreen,
urlcolor=indigo(dye),
}

\newcommand{\Mpl}{M_{\rm Pl}}

\newcommand{\e}{\epsilon}

\renewcommand{\O}{\mathcal{O}}
\newcommand{\Ha}{\mathcal{H}}
\renewcommand{\L}{\mathcal{L}}
\newcommand{\pa}{\partial}
\newcommand{\SF}{\rm SF}
\newcommand{\U}{\rm C}
\newcommand{\End}{\rm end}

\newcommand{\curv}{\zeta}

\def\fncb{$\overline{\mathrm{FNC}}$}

\DeclareMathOperator\Arg{Arg}

\newcommand*\diff{\mathop{}\!d}
\newcommand*\diffcubed{\mathop{}\!d^3}
\newcommand*\difffour{\mathop{}\!d^4}

\newcommand*\re[1]{\mathop{}\!\mathrm{Re}\left[#1 \right]}

\newcommand\numberthis{\addtocounter{equation}{1}\tag{\theequation}}
\definecolor{darkgreen}{cmyk}{0.85,0.2,1.00,0.2} 
\definecolor{purple}{cmyk}{0.5,1.0,0,0} 

\newcommand{\fNL}{f_{\rm NL}}
\newcommand{\Long}{{L}}
\newcommand{\Short}{{S}}
\newcommand{\Peak}{{\rm Peak}}
\newcommand{\kLong}{k_{\Long}}
\newcommand{\kShort}{k_{\Short}}
\newcommand{\kPeak}{k_{\Peak}}
\newcommand{\Squeezedks}{\kLong,\kShort,\kShort}
\newcommand{\USR}{\rm USR}
\newcommand{\SR}{\rm SR}

\newcommand{\Msun}{M_\odot}

\begin{document}

\preprint{YITP-18-128}

\title[]{Primordial Black Holes and Local Non-Gaussianity in Canonical Inflation}

\author{Samuel Passaglia}
\email{passaglia@uchicago.edu}
\affiliation{Kavli Institute for Cosmological Physics, Department of Astronomy \& Astrophysics, 
Enrico Fermi Institute, University of Chicago, Chicago, IL 60637}

\author{Wayne Hu}
\affiliation{Kavli Institute for Cosmological Physics, Department of Astronomy \& Astrophysics, 
Enrico Fermi Institute, University of Chicago, Chicago, IL 60637}

\author{Hayato Motohashi}
\affiliation{Center for Gravitational Physics, Yukawa Institute for Theoretical Physics, Kyoto University,
Kyoto 606-8502, Japan}

\label{firstpage}

\begin{abstract}
Primordial black holes (PBHs) cannot be produced abundantly enough to be the dark matter in canonical single-field  inflation under slow roll. This conclusion is robust to local non-Gaussian correlations between long- and short-wavelength curvature modes, which we show have no effect in slow roll on local primordial black hole abundances. For the prototypical model which evades this no go, ultra-slow roll (USR), these squeezed non-Gaussian correlations  have at most an order unity effect on the variance of PBH-producing curvature fluctuations for models that would otherwise
fail to form sufficient PBHs.  Moreover, the transition out of USR, which is necessary for a successful model, suppresses even this small enhancement  unless it causes a large increase in the inflaton kinetic energy in a fraction of an $e$-fold, which we call a large and fast transition. 
Along the way we apply the in-in formalism, the $\delta N$ formalism, and gauge transformations to compute non-Gaussianities and  illuminate different aspects of the physical origin of these results. Local non-Gaussianity in the squeezed limit does not weaken the Gaussian conclusion that PBHs as dark matter in canonical single-field inflation require a complicated and fine-tuned potential shape with an epoch where slow roll is transiently violated. 
\end{abstract}

\date{\today}

\maketitle 

\section{Introduction}

	Primordial black holes \cite{Zeldovich:1967aaa,Carr:1974,Carr:1975aaa,Meszaros:1974aaa,Chapline:1975aaa} (PBHs) can form in the early universe from the collapse upon horizon reentry of perturbations generated during inflation. PBHs of mass $\sim 10^{-11} \Msun$ could comprise the dark matter \cite{Carr:2016drx,Niikura:2017zjd,Carr:2017jsz,Kuhnel:2017pwq} if one can evade current astrophysics-dependent constraints from neutron-star capture in globular clusters \cite{Capela:2013yf,Lane:2009aa}, while PBHs of mass $\sim 10 \ \Msun$ could be responsible for LIGO black-hole merger events \cite{Bird:2016dcv,Sasaki:2016jop,Zumalacarregui:2017qqd,Sasaki:2018dmp,LIGOScientific:2018mvr}.

	PBHs must be formed on physical length scales far removed from the CMB and large-scale-structure modes, where the perturbations are too small, and therefore constraining their abundance and mass provides complementary information about the inflationary epoch. In particular, Ref.~\cite{Motohashi:2017kbs} showed that PBHs cannot be produced with sufficient abundance to be the dark matter through Gaussian fluctuations in canonical inflation, i.e.~by a single scalar field with a canonical kinetic term, without violating the slow-roll (SR) assumption.   Therefore any confirmed detection of PBHs as a significant mass fraction of the dark matter would provide evidence beyond canonical slow-roll inflation, e.g.~for a violation of slow roll after CMB modes exit the horizon (e.g., \cite{Ballesteros:2017fsr,Kawasaki:2016pql,Garcia-Bellido:2017mdw}), a non-canonical kinetic term for the inflaton (e.g., \cite{Kamenshchik:2018sig}), a multi-field inflationary scenario (e.g., \cite{Suyama:2011pu,Kawasaki:2012wr}), or some other non-standard cosmological scenario (e.g., \cite{Chen:2016kjx}).

	PBH abundances are sensitive to the full probability distribution of the density contrast averaged on horizon scales at reentry~\cite{Young:2013oia,Young:2014ana,Franciolini:2018vbk,Germani:2018jgr}. Since PBHs form from peaks of the density fluctuations, non-Gaussianity in the probability distribution can have a significant impact on predictions of PBH abundances. In this work we study whether non-Gaussianity can make a model that, under the Gaussian assumption, fails to form sufficient PBH dark matter into one that does, rather than how non-Gaussianity changes the abundance of a rare tracer population of PBHs.

	In canonical inflation, the predominant source of non-Gaussianity is of the local type, which is induced by local field or gauge transformation.  In the squeezed limit, the induced coupling of long- and short-wavelength modes generates a  local and position-dependent modulation of the power spectrum.  Here the local probability distribution of short-wavelength curvature perturbations remains Gaussian but the long-wavelength mode induces a global position-dependent modulation of the variance of the distribution.

	In \S\ref{sec:nogo} we show that the squeezed effect vanishes in freely-falling coordinates in slow roll, and therefore Ref.~\cite{Motohashi:2017kbs}'s no-go theorem for PBH formation in canonical inflation is robust to squeezed non-Gaussianity: any such model which produces large quantities of PBHs must break the slow-roll approximation well before the end of inflation.

	In \S\ref{sec:usr} we study the prototypical model which evades the no-go theorem by violating the
	slow-roll approximation -- the so-called {ultra-slow-roll (USR) inflation}~\cite{Kinney:2005vj} 
	-- to show analytically that even beyond slow roll, squeezed non-Gaussianity in canonical single-field inflation can have only an order unity effect on the local statistical properties of fluctuations for models which fail to form PBHs as the dark matter under Gaussian 
	assumptions. Along the way we show how the non-Gaussianity in USR can be understood intuitively via the $\delta N$ formalism \cite{Starobinsky:1985aa,Salopek:1990jq,Sasaki:1995aw,Sugiyama:2012tj,Domenech:2016zxn,Abolhasani:2018gyz}, and we discuss its relationship to locally observable quantities in freely-falling coordinates, specifically the local power spectrum that is relevant to PBHs. In App.~\ref{app:numerics} we review how local non-Gaussianities are computed numerically through the in-in formalism, and in App.~\ref{app:gauge} we clarify how this computation for
	USR is related through boundary terms to the analytic computation via gauge transformation, usually called a field redefinition in the literature \cite{Maldacena:2002vr}.

	Observationally viable implementations of USR inflation must have the USR phase be transient. In \S\ref{sec:transientUSR}, we explore the phenomenology of such transient USR models to show that in most circumstances the order-unity effect in pure USR is suppressed. In particular, in \S\ref{subsec:inflection} we study a transient USR model proposed in the literature for PBH formation, inflection-point inflation, and show that squeezed non-Gaussianities have a much smaller than order unity effect on the local power spectrum.
			
	To understand why the local power spectrum is not enhanced in the inflection-point model and to study generally the non-Gaussianity in transient USR, we first review in \S\ref{subsec:infinite} the analytic results of Ref.~\cite{Cai:2017bxr} for infinitely sharp potential transitions from USR to SR. We then generalize that study in \S\ref{subsec:flat} to different types of transitions from USR with the help of an adjustable toy model. 
	
	We show that transient USR recovers the results of pure USR only when the transition to SR causes  a large increase in the inflaton kinetic energy in a fraction of an $e$-fold, which we call a \textbf{large} and \textbf{fast} transition. Inflection-point inflation is, by contrast, an example of a \textbf{small} and \textbf{slow} transition.  Thus there is only a limited, but well defined, class of transient models that can have significant non-Gaussian changes to the local power spectrum of inflationary fluctuations due to squeezed correlations, and even these cannot enhance the formation of PBHs 
	sufficiently to be the dark matter unless the model is at least on the threshold of being able to
	produce them under Gaussian assumptions already.
	We conclude in \S\ref{sec:conclusion}.

\section{No Go for Slow Roll}
	\label{sec:nogo}

	Neglecting the non-Gaussianity of perturbations, Ref.~\cite{Motohashi:2017kbs} showed that in canonical single field inflation the comoving
	curvature power spectrum
	\begin{equation}
	\Delta_\curv^2  \equiv \frac{k^3}{2 \pi^2} P_\curv 
	\end{equation}
	must reach at least
	\begin{equation}
	\label{eq:PBHcriterion}
	\Delta^2_\curv \sim 10^{-2}
	\end{equation}
	within $\sim42$ $e$-folds from the epoch when CMB scales exited the horizon, at which $\Delta_\curv^2 \simeq
	10^{-9}$, for the dark matter to be entirely composed of PBHs. 
	In slow roll, the power spectrum satisfies
	\begin{equation}
	\label{eq:Delta2SR}
	\Delta^2_\curv\simeq
	\frac{H^2}{8 \pi^2 \e},
	\end{equation}
	with $\e \equiv - d \ln H / d N$, $N$ the negative increasing $e$-folds to the end of inflation, and the reduced Planck mass $\Mpl\equiv(8\pi G)^{-1/2}=1$ here and throughout. Therefore such an enhancement of $\Delta^2_\curv$ requires a slow-roll violation of at least $\eta \equiv d \ln \e / dN \sim  1$  after horizon exit of the CMB modes but well before the end of inflation.
	
	In this section we update this slow-roll no-go theorem to include local non-Gaussianity which modulates
	short-wavelength power in a long-wavelength mode. In particular 
	since the formation of a PBH depends on the density fluctuation averaged on the horizon
	scale at 
	reentry of the perturbations, horizon scale power that is modulated by superhorizon wavelength fluctuations can
	in principle enhance formation.  We study whether such a modulation can make it possible to produce a 
	substantial fraction of the dark matter in PBHs with slow-roll inflation.

	In the presence of a long-wavelength fluctuation $\curv_{\Long}$, low pass filtered for 
	 comoving wavenumbers $k\le k_L$, the power spectrum at $\kShort \gg \kLong$ becomes position dependent
	\begin{equation}
	\label{eq:modulation}
	P_\curv(\kShort, x) = P_\curv(\kShort) \left[ 1 + \frac{d \ln P_\curv(\kShort)}{d \curv_{\Long}} \curv_{\Long} (x) \right].
	\end{equation}
	 By multiplying by and averaging over the long-wavelength mode, 
	\begin{eqnarray}
	\langle \curv_{\Long}(x) P_\curv(\kShort, x) \rangle_{\curv_{\Long}} &\simeq &  \int\frac{d^3 k_L}{(2\pi)^3}
	B_\zeta(\kLong,\kShort,\kShort),
	\end{eqnarray}
	we can relate the power spectrum response to the curvature bispectrum
	 $B_\zeta$,
	 \begin{equation}
	\label{eq:response}
	\frac{d \ln P_\curv(\kShort)}{d \curv_{\Long}} \simeq \frac{	B_\zeta(\kLong,\kShort,\kShort)} {P_\curv(\kShort) P_\curv(\kLong)} \simeq \frac{12}{5} \fNL(\Squeezedks).
	\end{equation}
	Here $\fNL$ is the standard dimensionless non-Gaussianity parameter
	\begin{equation}
	\label{eq:fNLdefinition}
	\fNL (k_1, k_2, k_3) \equiv \frac{5}{6} \frac{B_\curv(k_1, k_2, k_3)}{P_\curv(k_1) P_\curv(k_2) + \text{perm.}} ,\ \ 
	\end{equation}
	in which `$+\text{ perm.}$' denotes the two additional cyclic permutations of indices and the
	approximation \eqref{eq:response} assumes the squeezed limit $\kShort \gg \kLong$.

	\begin{figure*}[t]
	\psfig{file=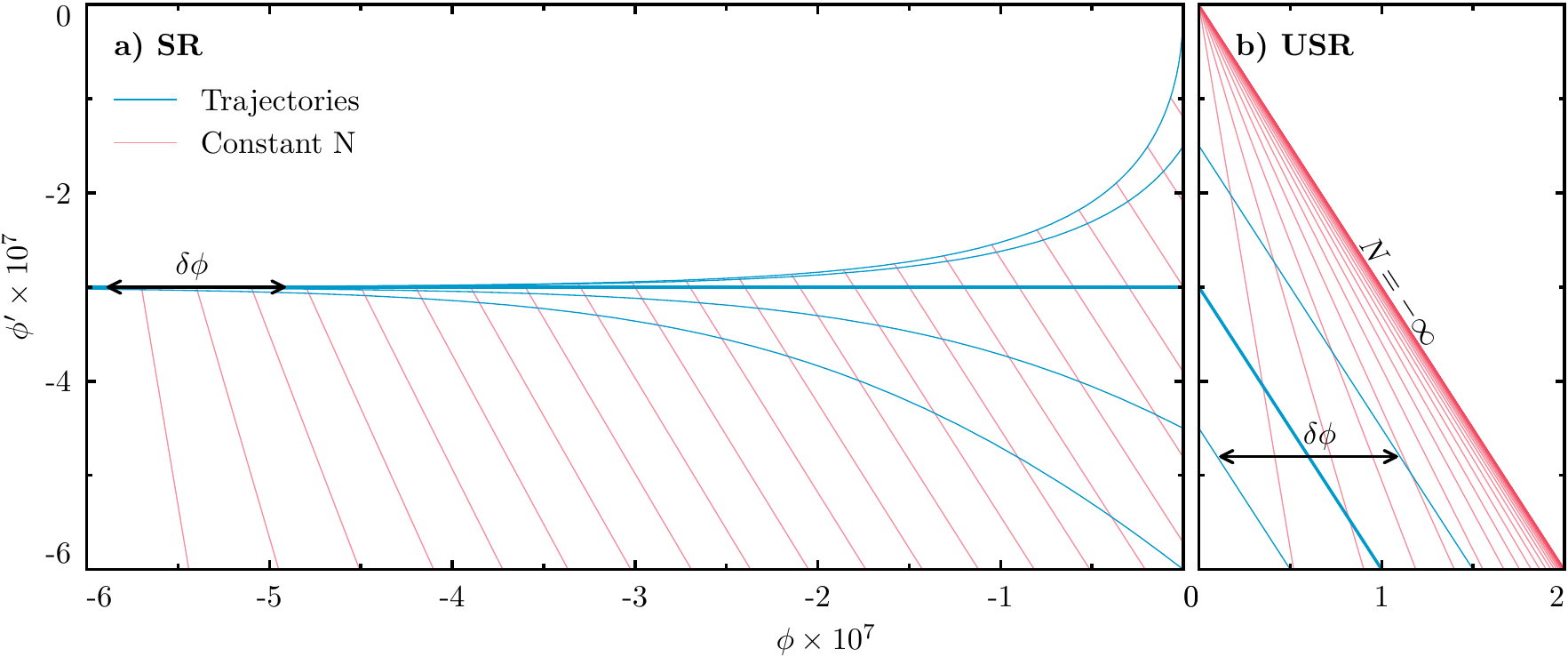}
	\caption{Phase space diagram for a) slow roll (SR) and b) ultra-slow roll (USR). Shown are
	background trajectories (blue lines), lines of constant $e$-folds (red lines) to the end of inflation (left edge of panels), and field fluctuations $\delta \phi$ (arrows).
	SR trajectories converge to the attractor for different initial kinetic energies at $\phi=0$.  SR field fluctuations
	$\delta \phi$  follow the  attractor trajectory and can be absorbed into a change in  $e$-folds leaving a change in the relationship between local and global coordinates, but no local imprint on observables  once clocks are synchronized to a fixed field value at the end of inflation. 	
	USR field fluctuations  can still be absorbed into a local background but no longer the background
	of the unperturbed universe (thick blue line).   Since different USR trajectories experience  different numbers of $e$-folds to the end,
	the power spectrum becomes position dependent, with $\fNL$ reflecting the $e$-folding asymmetry between
	positive and negative $\delta\phi$ or $\partial^2 N/\partial \phi^2$. 
		} 
	\label{fig:phasespaceSR+USR}
	\end{figure*}

	In single-field inflation, $\fNL(\Squeezedks)$ has a constrained form when $\zeta$ is conserved above the horizon. The curvature perturbation is equivalent to a field fluctuation in spatially flat
	gauge $\zeta=-\delta\phi/\phi'$, with primes denoting derivatives with respect to 
	$e$-folds  $'=d/dN$ here and throughout.  Therefore for a constant $\zeta$, the field fluctuation evolves according to 
	\begin{equation}
	 \delta\phi' = \frac{\phi''}{\phi'}\delta\phi,
	 \label{eq:attractor}
	\end{equation}
	and the phase-space trajectory of the long-wavelength field perturbation follows that of the background
	itself.  Short-wavelength modes evolving in a long-wavelength perturbation then also follow the phase-space trajectory of the background, with the only difference being the local $e$-folds which determines the relationship between physical and comoving wavenumber (see Fig~\ref{fig:phasespaceSR+USR}\hyperref[fig:phasespaceSR+USR]{a}). 

	Single-field inflation on the slow-roll attractor \eqref{eq:attractor} therefore satisfies the consistency relation \cite{Maldacena:2002vr}
	\begin{equation}
	\lim_{\kLong/\kShort\rightarrow0}\frac{12}{5} \fNL (\Squeezedks) = -\frac{d \ln{ \Delta_\curv^2(\kShort)}}{d \ln{\kShort}}.
	\label{eq:ConsistencyRelation}
	\end{equation}

	This implies a modulation of the small-scale power spectrum due to the long-wavelength mode according to Eqs.~\eqref{eq:modulation} and~\eqref{eq:response} as
	\begin{equation}
	\label{eq:GlobalDilation}
	P_\curv(\kShort, x) = P_\curv(\kShort) \left[ 1 -\frac{d \ln{ \Delta_\curv^2(\kShort)}}{d \ln{\kShort}} \curv_{\Long} (x) \right].
	\end{equation}
	This modulation is zero at the scale where the power spectrum peaks and corresponds to a dilation of scales rather than an amplitude enhancement.  
	
	In general the physical effect of a  dilation of scales is to change the mass scale of PBHs rather than enhance their abundance.  However in the slow-roll case, there is actually a change in neither abundance nor mass scale.
	Though the dilation \eqref{eq:GlobalDilation} does occur in global comoving coordinates, in single-field inflation a freely-falling
	observer will not see this dilation locally. 

	For a given perturbed metric, 	the standard Fermi normal coordinates (FNC)~\cite{Manasse:1963} can be constructed with respect to  a central timelike geodesic of a comoving observer~\cite{Senatore:2012ya,Senatore:2012wy}, such that $g_{\mu\nu}^{\rm FNC} \simeq \eta_{\mu\nu}$ up to tidal corrections.
	In order to absorb the effects of superhorizon perturbations out to the horizon scale of a local observer, as required for PBH calculations, we utilize  conformal Fermi normal coordinates (\fncb{}) \cite{Pajer:2013ana}.  \fncb{} are
	constructed such that $g_{\mu\nu}^{\overline {\rm FNC}} \simeq a^2 \eta_{\mu\nu}$, i.e.~a conformally flat, locally 
	Friedmann-Lema\^itre-Robertson-Walker (FLRW) form where the global scale factor $a$ of the background
	universe is evaluated
	at the proper time of the central observer.
	
	As shown in Ref.~\cite{Pajer:2013ana}, for single-field slow-roll inflation the bispectrum in \fncb{}
	is related to the comoving-gauge bispectrum by an additional term proportional to the tilt of the power spectrum as 
	\begin{align*}
	\label{eq:Bzetatrans}
	\lim_{\kLong/\kShort\rightarrow0} B_{\bar\curv}(\Squeezedks) =\:& P_\curv(\kLong) P_\curv(\kShort) \frac{d\ln \Delta^2_\curv(\kShort)}{d\ln \kShort} \\
	& + B_{\curv}(\Squeezedks),
	\numberthis
	\end{align*}
	where barred symbols denote quantities in the \fncb{} frame. This additional term neatly cancels the comoving-gauge squeezed bispectrum from the consistency relation~\eqref{eq:ConsistencyRelation} 
	and thus in single-field slow-roll inflation
	\begin{equation}
	\label{eq:BzetabarCancellation}
	\lim_{\kLong/\kShort\rightarrow0}  B_{\bar\curv}(\Squeezedks) = 0.
	\end{equation}
	There is therefore no modulation of the power spectrum in \fncb{} 
	\begin{equation}
	\label{eq:PzetabarModulation}
	P_{\bar\curv}(\kShort, x) = P_{\bar\curv}(\kShort),
	\end{equation}
	and the small-scale power spectrum in \fncb{} does not depend on the value of the long-wavelength perturbation. All local observers therefore see the same small-scale power spectrum regardless of their position in the long-wavelength mode. 
	
	Physically, the cancellation in Eq.~\eqref{eq:BzetabarCancellation} occurs because the bispectrum from the consistency relation encodes the effect on small-wavelength modes of evolving in a separate universe with a background evolution defined by the long-wavelength mode. Once the long-wavelength mode is frozen, this effect is just to change coordinates in the separate universe relative to global coordinates. When making local observations, an observer knows nothing of the global coordinates and instead makes measurements in coordinates corresponding to the separate universe. The formation of
	PBHs is a local process and so their properties also do not depend on their position in the long-wavelength
	mode.

	This lack of local modulation can also be understood from the phase-space diagram Fig.~\ref{fig:phasespaceSR+USR}\hyperref[fig:phasespaceSR+USR]{a}. 
	Relative to the end of inflation 
	at a fixed field value, perturbed trajectories in slow roll are indistinguishable from the background trajectory and thus observers making measurements relative to the end of inflation
	cannot from any local measurement decide whether they inhabit different regions of a long-wavelength
	curvature perturbation.
	
	This leads us to our first conclusion:  squeezed non-Gaussianity cannot produce PBHs as a significant fraction of the dark matter in canonical single-field slow-roll inflation.  For such PBHs to form in canonical
	single-field inflation, the slow-roll approximation must be violated, at least transiently, to either produce large Gaussian
	or non-Gaussian fluctuations. In this sense, the slow-roll no-go theorem shown in Ref.~\cite{Motohashi:2017kbs} is robust and does not change.

	Models that evade this no-go result typically have a period when the inflaton rolls on a very
	flat potential where Hubble friction is insufficient to keep the inflation on the slow-roll attractor.  
	The ultra-slow-roll model, where the inflaton potential is perfectly flat, provides the prototypical example for
	such studies as we shall  see next.

\section{Ultra-Slow-Roll Inflation}
	\label{sec:usr}

	Ultra-slow roll~\cite{Kinney:2005vj} is a model of single-field inflation which greatly enhances the scalar  power spectrum while also breaking the single-field consistency relation \eqref{eq:ConsistencyRelation}  for the squeezed bispectrum by violating the attractor condition (\ref{eq:attractor})~\cite{Namjoo:2012aa,Martin:2012pe}. It is therefore possible to spatially modulate the local power in small scale
	density fluctuations relevant for PBHs with long-wavelength modes.
	In this section we examine whether this non-Gaussian modulation can significantly enhance the PBH abundance in ultra-slow roll.

	USR is characterized by a potential which is sufficiently flat before its end, which we denote with $\phi=0$, that the Klein-Gordon equation takes the form 
	\begin{equation}
	\ddot\phi \simeq - 3 H \dot\phi,
	\label{eq:KGUSR}
	\end{equation}
	where here and throughout overdots denote derivatives with respect to the coordinate time $t$. If the
	potential energy dominates then $H\simeq \text{const.}$ and	Eq.~\eqref{eq:KGUSR} then implies $\phi''\simeq -3\phi'$ and hence $\phi'\simeq -3\phi+\text{const.}$, defining a family of trajectories in the phase-space diagram, as depicted by the blue trajectories in Fig.~\ref{fig:phasespaceSR+USR}\hyperref[fig:phasespaceSR+USR]{b}.  
	Therefore, the phase-space trajectory of the background evolution depends on the initial kinetic energy and does not exhibit attractor behavior.

	For an exactly flat potential at $\phi>0$, an inflaton with insufficient initial kinetic energy will not cross the plateau to 
	reach $\phi=0$, neglecting stochastic effects.
	In Fig.~\ref{fig:phasespaceSR+USR}\hyperref[fig:phasespaceSR+USR]{b} we focus on classical trajectories that can reach $\phi=0$ within finite $e$-folds, and hence the upper right triangle region is inaccessible.

	The solution to Eq.~\eqref{eq:KGUSR} is $\dot\phi \propto a^{-3}$ and so $\e\propto a^{-6}$ and $\eta = -6$.   
	Since the analytic solution of the Mukhanov-Sasaki equation for $\curv$ in the superhorizon limit is given by
	\begin{equation}
	\label{eq:shsol-s} 
	\curv \simeq c_1 + c_2 \int \frac{\diff{t}}{a^3 \e} , 
	\end{equation}
	with integration constants $c_1$ and $c_2$, it is dominated by the second mode which 
	grows in USR since $(a^3 \e)^{-1} \propto a^{3}$ rather than decays as it does in slow roll.
	With $H\simeq \text{const.}$, Eq.~\eqref{eq:shsol-s} gives $\zeta\propto a^{3}$ and hence in the spatially flat gauge $\delta\phi=-\zeta\phi'=\text{const.}$, implying that $\delta\phi'=0$, unlike the case of the slow-roll attractor \eqref{eq:attractor}.

	The power spectrum in this model depends on the value of $\e$ at the end of USR [see Eq.~\eqref{eq:USRmodefunction}],
	\begin{equation}
	\label{eq:Delta2USR}
	\Delta^2_\curv\simeq
	\dfrac{H^2}{8 \pi^2 \e_{\text{end}}},
	\end{equation}
	and thus can be very large if $\e_{\End} \ll 1$.  	One can employ a gauge transformation from spatially flat gauge to comoving gauge to show that  the squeezed-limit non-Gaussianity takes the form (see Appendix~\ref{app:gauge})
	\begin{equation}
	\label{eq:USRfNL}
	\lim_{\kLong/\kShort\rightarrow0}\frac{12}{5} \fNL (\Squeezedks) = 6.
	\end{equation}
		Since the USR power spectrum is scale invariant,  the large value of $\fNL$ in USR violates
	the consistency relation. 

	The physical origin of this large value for $\fNL$ can  be seen from the phase-space diagram Fig.~\ref{fig:phasespaceSR+USR}\hyperref[fig:phasespaceSR+USR]{b}.
	Due to the initial kinetic energy dependence of the background evolution, a USR perturbation cannot be mapped into a change in the background clock along the same phase-space trajectory. Instead, long-wavelength perturbations $\delta\phi$ carry no corresponding $\delta\phi'$ and so shift the USR trajectory to one with a different relationship between $\phi$ and $\phi'$.   On this shifted trajectory, the short-wavelength power spectrum attains a different value at the end of USR.  More generally, if a local measurement is sensitive to $\phi'$ at the end of inflation, as in the case of $\Delta^2_\curv(\kShort)$, then different observers will produce different measurements depending on their position in the long-wavelength mode. 
	
	This graphical representation of $\fNL$ can be turned into a computational method through
	the so-called $\delta N$ formalism  \cite{Starobinsky:1985aa,Salopek:1990jq,Sasaki:1995aw,Sugiyama:2012tj}. When the expansion shear for a local observer is negligible, as it is in USR above the horizon, the nonlinear evolution of the curvature fluctuation
	follows the evolution of local $e$-folds.    On spatially flat hypersurfaces, the field fluctuation can
	be absorbed into a new conformally flat FLRW background on scales much shorter than the
	wavelength and so the local $e$-folds may be calculated from the Friedmann equation of a separate
	universe.  The position-dependent power spectrum is therefore the second order change in $e$-folds due to 
	a short-wavelength $\delta \phi_S$ on top of a long-wavelength $\delta \phi_L$.  Since in USR these perturbations
	leave $\phi'$ unchanged, the non-Gaussianity parameter can be computed from the $e$-folds as a function of 
	phase-space position of the background $N(\phi,\phi')$  as
	\begin{equation}
	\label{eq:deltaN}
	\frac{12}{5} \fNL = 2 \frac{\partial^2 N}{\partial \phi^2} \left/ \left(\frac{\partial N}{\partial \phi}\right)^2 \right. ,
	\end{equation}
	at fixed $\phi'$. 	
	
	The consequence of this formula can be visualized through Fig.~\ref{fig:phasespaceSR+USR}\hyperref[fig:phasespaceSR+USR]{b}
	 as the effect of perturbations on phase-space trajectories. Around a chosen background trajectory, the long-wavelength perturbation is reabsorbed into a new background, a horizontal shift to a new trajectory. Short-wavelength perturbations living in this new background induce a second shift in the trajectory, hence the second derivative.  Visually, the fact that for the same amplitude of field fluctuation $|\delta\phi|$, a positive fluctuation intersects more surfaces of constant $N$ than
	a negative fluctuation indicates a large $\fNL$.
	Refs.~\cite{Namjoo:2012aa,Chen:2013eea,Cai:2017bxr,Pattison:2017mbe} follow this approach to analytically 
	compute its value in complete agreement with the in-in approach App.~\ref{app:numerics} or the gauge-transformation App.~\ref{app:gauge}. We shall again exploit the $\delta N$ formalism in \S\ref{sec:transientUSR}.

	Despite the violation in the consistency relation, the coordinate transformation for the bispectrum Eq.~\eqref{eq:Bzetatrans} still holds and the transformation from global comoving coordinates to \fncb{} leads to the same additional tilt-dependent term in the bispectrum as in the canonical case so long as the transformation to \fncb{} is performed when modes are frozen outside the horizon after the end of inflation.\footnote{
	\fncb{} can still be established during the USR phase but are more closely related
	to spatially flat gauge than comoving gauge in temporal synchronization (see also App.~\ref{app:gauge}).  In spatially flat gauge, a superhorizon field fluctuation $\delta\phi$ can 
	be absorbed into a new, nearly conformally flat FLRW background, as we  exploit with the 
	$\delta N$ formalism.}
	After this time, the construction follows Ref.~\cite{Pajer:2013ana} exactly. This procedure of transforming coordinate systems after inflation is followed for slow-roll inflation in Ref.~\cite{Cabass:2016cgp} to compute the next-to-leading order term in the bispectrum transformation. Practically, it corresponds to the clock-synchronization condition that all local observers make their measurements at fixed proper time after the end of inflation.
	 
	 Given the scale invariance of the spectrum, the tilt-dependent transformation from comoving gauge to \fncb{} 
	leaves neither an enhancement of the local power in the long-wavelength mode nor 
	a modulation of the mass of the PBHs.  On the other hand, since the transformation term no longer cancels with the comoving-gauge $\fNL$ 
	itself, a large value of the latter can in principle enhance PBH formation locally. 

	If $\fNL(\Squeezedks)$ is described by the USR result Eq.~\eqref{eq:USRfNL}, then the local power spectrum can be enhanced by a factor $12/5\times \fNL \times \curv_{\Long} = 6 \times \curv_\Long$. Therefore the non-Gaussian response enhances the local power spectrum by an order unity quantity unless the long-wavelength mode is large, i.e.\ 
	\begin{equation}
	\curv_\Long \gtrsim 10^{-1}.
	\end{equation}
	However, the scale invariance of USR would then imply
	\begin{equation}
	\Delta^2_\Short = \Delta^2_\Long \sim \langle \curv_\Long^2(x)\rangle \gtrsim 10^{-2},
	\end{equation}
	which satisfies the criterion Eq.~\eqref{eq:PBHcriterion} for PBH formation, and therefore PBHs would already be produced at scale $\kShort$ even before accounting for the non-Gaussian response. Note that the conversion from $\Delta^2_\curv$ to spatial variance involves a summation over $\kLong$ and
	gives a logarithmic factor which depends on the total $e$-folds of USR.   In a realistic model this logarithmic 
	factor must be finite 
	so as to also satisfy constraints from the CMB.   

	This result is the second main conclusion of this work: in a USR model which does not produce a significant PBH abundance under the Gaussian approximation, the squeezed non-Gaussian response enhances the local power spectrum by at most
	\begin{equation}
	 \frac{\Delta P_\curv}{P_\curv} \lesssim 1,
	\end{equation}
	and therefore the squeezed non-Gaussian response does not qualitatively change Gaussian conclusions.  
	Of course as they originate from rare fluctuations, PBHs can change in their abundance but these changes can 
	be reabsorbed into model parameters that make no more than an order unity change in the power spectrum.
	In particular squeezed non-Gaussianity cannot make a model that falls far short of making PBHs the dark matter under
	the Gaussian assumption into one that does.

	Since inflation has to end and observational constraints should be satisfied on CMB scales, the simple picture presented here must be modified to account for transitions into and out of USR.   In \S\ref{sec:transientUSR} we shall explore whether even this level of enhancement still holds in such models of transient USR inflation.

\section{Transient Ultra-Slow Roll}
	\label{sec:transientUSR}
	
	In addition to a graceful exit problem, USR inflation is incompatible with the measured tilt of the CMB power spectrum \cite{Aghanim:2018eyx} and is in tension with constraints on local non-Gaussianities in the CMB \cite{Ade:2015ava}, and therefore any USR phase must begin after CMB modes exit the horizon and must take care not to grow those modes after horizon exit.

	One model proposed in the literature for PBH production with a transient USR phase is inflection-point inflation \cite{Motohashi:2017kbs,Garcia-Bellido:2017mdw}. In \S\ref{subsec:inflection}, we show that the transition out of USR in inflection-point inflation induces
	\begin{equation}
	\fNL(\Squeezedks) \ll 1,
	\end{equation}
	and therefore non-Gaussianities do not enable  PBHs to be the dark matter in inflection-point inflation.

	This numerical result can be understood from Ref.~\cite{Cai:2017bxr}'s analytic study of infinitely sharp potential transitions between USR and SR, which we review briefly in \S\ref{subsec:infinite}.  Transitions where the inflaton velocity monotonically decreases to reach an attractor solution lead to squeezed non-Gaussianity that is proportional to the potential slow-roll parameters on the attractor. Conversely, transitions where the inflaton instantly goes from having too much kinetic energy for the potential it evolves on to suddenly having insufficient kinetic energy for a now much steeper potential conserve the USR non-Gaussianity.  We call the latter transitions
	 \textbf{large}, which we will define specifically below [see Eq.~\eqref{eq:CaiCriterion}].  

	In \S\ref{subsec:flat}, we generalize the analysis of Ref.~\cite{Cai:2017bxr} to potentials which do not have an infinitely sharp break, and in particular we study how quickly the inflaton must traverse the potential feature to reproduce the USR result. We show that to conserve the USR result Eq.~\eqref{eq:USRfNL} the transition must be \textbf{fast} in that it completes in a small fraction of an $e$-fold [see Eq.~\eqref{eq:dN}].   

	We conclude that the large $\fNL$ of USR will only be preserved if the transition to SR is both
	large and fast. For all other cases,  the enhancement to the local power spectrum
		\begin{equation}
	\frac{\Delta P_\curv}{P_\curv} \ll 1,
	\end{equation}
	and so squeezed non-Gaussianity in transient USR does not generally affect the conclusions on PBH formation.

	\subsection{Slow--Small Transition: Inflection-Point Inflation}
		\label{subsec:inflection}

		Inflection-point inflation is characterized by a potential which supports a slow-roll phase when CMB scales exit the horizon followed by a slow-roll violation and subsequent ultra-slow-roll phase which enhances the power spectrum at small scales. This USR phase is generally unstable and lasts just a few $e$-folds before the inflaton loses enough kinetic energy to lock onto the attractor solution of the potential and slow-roll inflation resumes \cite{Pattison:2018bct}.	We call this transition \textbf{slow} because the inflaton kinetic energy decreases monotonically to the slow-roll value, and \textbf{small} because the potential slow-roll parameters on the attractor are comparable to the kinetic energy at the end of the USR phase.

		We consider an inflection potential of the form explored in Ref.~\cite{Motohashi:2017kbs} following Ref.~\cite{Garcia-Bellido:2017mdw},
		\begin{equation} 
			V(\phi) = \frac{\lambda v^4}{12} \frac{x^2 (6 - 4 a x + 3 x^2)}{(1 + b x^2)^2},
			\label{eq:GBInflectionPotential}
		\end{equation}
		where $x=\phi/v$. We study this model with the parameters $\left\{a, b-1, \lambda, v\right\} = \left\{3/2,\ 4 \times 10^{-5} ,\ 7 \times 10^{-8},\ 0.658 \right\}$. In terms of the auxiliary variables of Refs.~\cite{Motohashi:2017kbs,Garcia-Bellido:2017mdw}, this model has $\left\{\beta,\ \Delta N_{\SR} \right\}=\left\{4\times10^{-5},\ 125\right\}$.
	
		These parameters are finely tuned to significantly suppress $\e$ after the CMB scale $k_0 = 0.05\ \text{Mpc}^{-1}$ exits the horizon $55$ $e$-folds before the end of inflation while also preventing the inflection point from trapping the inflaton for too many $e$-folds. Nonetheless our qualitative results for the non-Gaussianity are not sensitive to the specific functional form of the potential nor to the parameter set above.  

		Note that even with fine-tuning, this model does not fit observational constraints from the CMB	(e.g.,~\cite{Aghanim:2018eyx}) because the power spectrum is too red (scalar slope $n_s = 0.91$) due to the proximity of the inflection point to CMB scales.	This additional red tilt implies a larger value of $\epsilon$ at CMB scales and hence a larger relative suppression of
		$\epsilon$ and growth of the power spectrum during the USR phase.   Without this enhancement, the
		inflection model falls far short of forming PBHs as the dark matter~\cite{Motohashi:2017kbs} and so we 
		choose these parameters to study whether models on the threshold of forming sufficient PBHs for Gaussian 
		fluctuations can be made to do so through non-Gaussianity in the model.

		\begin{figure}[t]
		\psfig{file=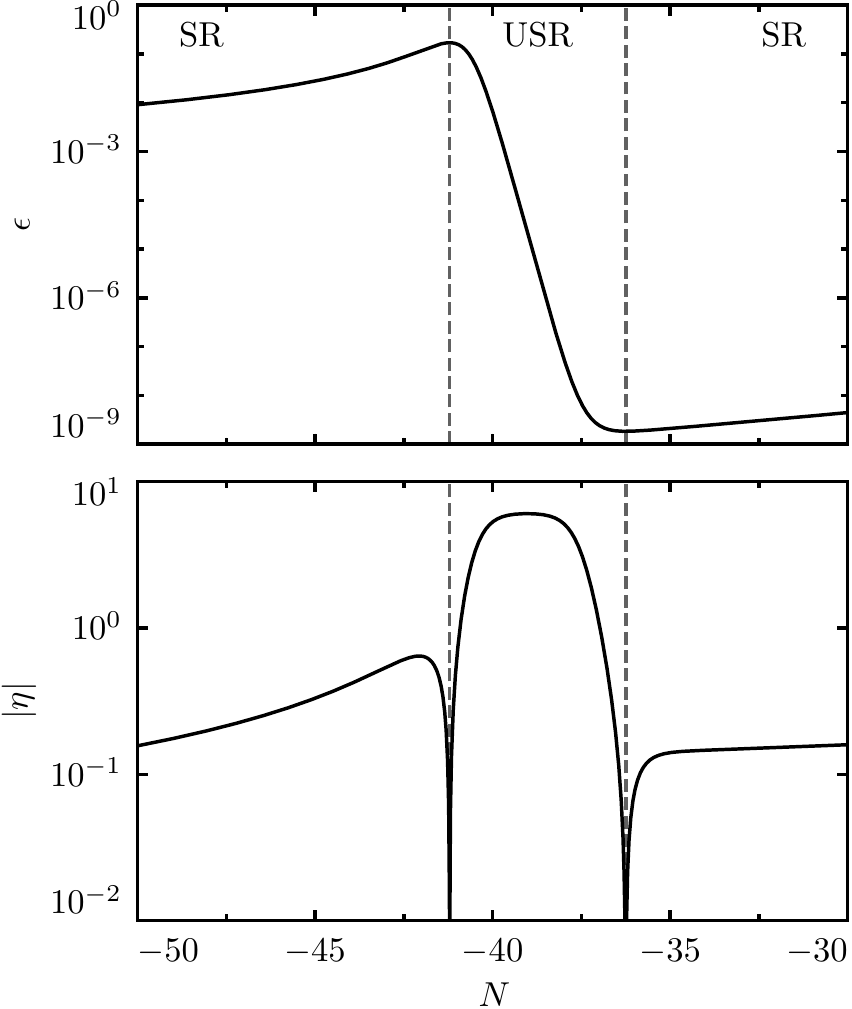}
		\caption{Inflection-point model background parameters $\e \equiv - d\ln H/d N$ and $\eta \equiv d\ln\e / d N$. $\eta$ experiences two zero-crossings as $\epsilon$ reaches critical points entering and exiting the USR period, which we use to delineate the USR phase from the SR phases.
				Here the transient period lasts for $\sim5$ $e$-folds, but only achieves $\eta \simeq -6$ for 
		a shorter period. }
		\label{fig:GBBackground}
		\end{figure}

		Along the inflaton trajectory, the potential~\eqref{eq:GBInflectionPotential} has a single inflection point, where $d^2V/d\phi^2 = 0$ is satisfied, between two close points where $dV/d\phi = 0$.  In this region the slope of the potential is tiny, and hence the USR condition $|dV/d\phi| \ll |\phi'| H^2$ is satisfied briefly, after which slow roll quickly resumes.
		The evolution of the slow-roll parameters $\e$ and $\eta$ in this model is shown in Fig.~\ref{fig:GBBackground}. The model exhibits a transient period where $\e \propto a^{-6}$ and thus the USR result $\eta\simeq-6$ is temporarily achieved. 

		\begin{figure}[t]
		\psfig{file=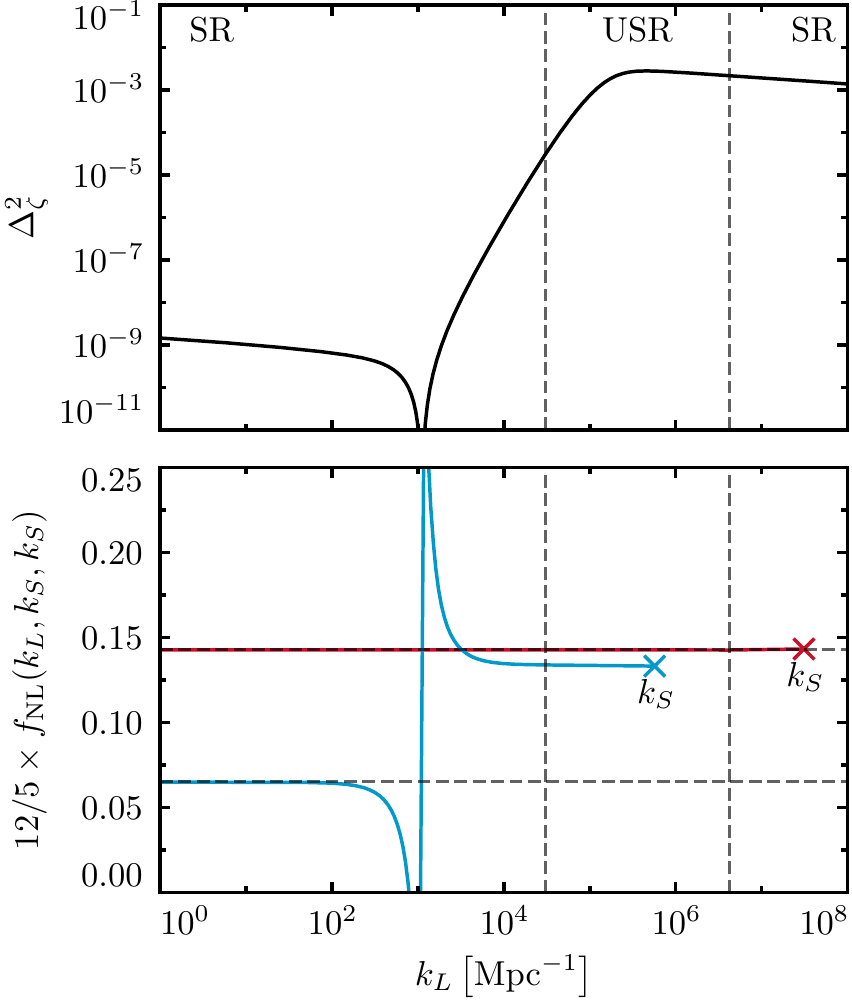}
		\caption{ Inflection-point model power spectrum $\Delta^2_\curv$ and  non-Gaussianity parameter $\fNL(\Squeezedks)$. The vertical dashed lines delineate modes which cross the horizon during the SR and USR periods (see Fig.~\ref{fig:GBBackground}). 
		For $\fNL$, $\times$'s denote values for $\kShort>\kLong$ with blue a mode that exits the horizon during USR and red after USR. 
		The horizontal dashed lines denote the consistency relation expectation for the two modes and the spike 
		in $\fNL$ reflects the near zero in $\Delta_\zeta^2$ rather than a large bispectrum
		 (see \S\ref{subsec:inflection} for further discussion).
		 }
		\label{fig:GBPowerSpecAndfNL}
		\end{figure}

		The upper panel of Fig.~\ref{fig:GBPowerSpecAndfNL} shows the power spectrum $\Delta^2_\curv$ produced by inflection-point inflation with the potential~\eqref{eq:GBInflectionPotential}, computed by numerically solving the Mukhanov-Sasaki equation \eqref{eq:MukhanovSasaki}. 
		Modes which exit the horizon well before the USR phase do not grow outside the horizon and their power spectrum satisfies the slow-roll result \eqref{eq:Delta2SR}. Modes which exit the horizon shortly before the USR phase do, however, grow outside the horizon leading to the rise before peak power in the USR phase.

		This behavior can be understood in more detail from the exact, but formal, solution of the Mukhanov-Sasaki equation	\eqref{eq:MukhanovSasaki} \cite{Hu:2016wfa}
		\begin{equation}
		\label{eqn:Rprimesoln}
		\curv'= -\frac{1}{a^3 \e H}\biggl[  \int \frac{d a}{a} a^3 \left(\frac{k}{aH}\right)^2 (\e H) {\curv} +{\rm const.}\biggr]. 
		\end{equation}
		In the SR phase, $\e$ is roughly constant and the growing integrand provides the leading contribution 
		\begin{equation}
		\label{eqn:RprimesolnSR}
		\curv'\simeq -\left(\frac{k}{aH}\right)^2  {\curv} ,  \qquad {\rm (SR)},
		\end{equation}
		and hence the curvature perturbation $\curv \propto e^{\frac{1}{2}\left(\frac{k}{aH}\right)^2}$ freezes out to a constant for $k/(aH)\ll 1$ as in Eq.~\eqref{eq:shsol-s}.
		On the other hand, in the USR phase, since $\zeta \propto a^3$ outside the horizon from \eqref{eq:shsol-s}, it immediately holds that
		\begin{equation} 
		\label{eqn:RprimesolnUSR} \curv' \propto a^3,  \qquad {\rm (USR)},
		\end{equation}
		outside the horizon. 
		One can also see that \eqref{eqn:RprimesolnUSR} is consistent with \eqref{eqn:Rprimesoln} as follows.  
		With $\zeta \propto a^3$, the integral in \eqref{eqn:Rprimesoln} acts as 
		$\simeq \int d \ln a \, a^{-2}$, which is dominated by early times and hence converges to a constant, whereas
		the prefactor grows as $\propto a^3$, resulting in \eqref{eqn:RprimesolnUSR}.
		Thus for a mode which spends $N_{\SR}$ $e$-folds outside the horizon in slow roll, it takes $N_{\USR} = 2/3 \times N_{\SR}$ $e$-folds of USR inflation to raise ${\curv}'$ back to order unity. Therefore at a fixed duration  $N_{\USR}$ of USR inflation, modes which exit the horizon more than $3/2 \times N_{\USR}$ $e$-folds before USR remain constant while modes which exit within $3/2 \times N_{\USR}$ grow outside the horizon. 
		After the USR phase, modes freeze in and the smooth change in the slope of the potential assures
		a slow increase in $\epsilon$ and a smooth transition of the power spectrum to the final SR phase.

		The power spectrum shown in the upper panel of Fig.~\ref{fig:GBPowerSpecAndfNL} exhibits a near-zero minimum $\Delta^2_\curv \sim 8 \times 10^{-16}$.
		Similar behavior occurs in other models in which the growing mode overtakes the constant mode (see, e.g., Refs.~\cite{Cicoli:2018asa,Ozsoy:2018flq,Byrnes:2018txb} and \S\ref{subsec:flat}). 
		This phenomenon can also be understood in detail from the formal solution Eq.~\eqref{eqn:Rprimesoln}, in which it can be seen that in slow roll the superhorizon mode approaches its slow-roll freezeout value with decreasing amplitude, i.e.\ with
		\begin{equation}
		\label{eq:FreezeOutArg}
		\Arg\left[\frac{\curv'}{\curv}\right] = \pi  + \O\left(\frac{k}{a H}\right) , \qquad {\rm (SR)},
		\end{equation}
		where the order of the correction follows from using the approximate SR form \eqref{eq:BunchDavies}
		in \eqref{eqn:Rprimesoln}.

		While $\zeta\propto a^3$ in the USR superhorizon limit, at the onset of USR, the curvature perturbation must reach this limit from the SR side.     Let $a=a_*$ at the onset of the USR phase, then the curvature
		evolves as
		\begin{equation}
		\curv' = \curv' \Big\vert_{a_*}  \times \left(\frac{a}{a_{*}}\right)^3 , \qquad {\rm (USR)},
		\end{equation}
		with the boundary condition $\curv' \vert_{a_*}$ given approximately by the SR solution for a smooth transition. Given the relative sign in the leading order SR expression \eqref{eqn:RprimesolnSR}, this represents an increase in the decay rate of $|\zeta|$ and thus before modes can grow as $\zeta \propto a^3$ in USR they must reverse sign. 

		There is a mode which experiences just enough evolution outside the horizon by the end of the USR phase to go from its freezeout value to near-zero.   The corresponding value of the power spectrum at the minimum is determined by the small out-of-phase component, i.e.\ how close Eq.~\eqref{eq:FreezeOutArg} is to $\pi$ and therefore how far outside the horizon this mode is when USR begins. Thus the longer the USR phase is, the deeper the minimum is.

		Modes which exit after this minimum are dominated by their superhorizon growth, and as modes exit the horizon closer to the USR phase they grow for a longer period and thus the power spectrum grows with increasing $k$. While a prolonged USR phase leads to a constant  $\Delta^2_\curv$ for modes which exit the horizon during USR [cf. Eq.~\eqref{eq:Delta2USR}], the inflection model touches the $\eta\simeq-6$ phase only briefly and the power spectrum therefore exhibits a peak  $\Delta^2_{\curv} (\kPeak) = 2.8 \times 10^{-3}$. This peak falls a factor of a few short of the value $\Delta^2_\curv\sim10^{-2}$ required for PBHs to form all the dark matter (see \S\ref{sec:nogo} and note that a model with the right tilt at CMB scales must fall much further short of this requirement \cite{Motohashi:2017kbs}). After the USR phase, the model returns to the slow-roll attractor and $\Delta^2_\curv$ is once more described by Eq.~\eqref{eq:Delta2SR}.

		It is now interesting to ask whether the power spectrum of the upper panel of Fig.~\ref{fig:GBPowerSpecAndfNL} can be locally enhanced by a factor of a few to exceed the threshold~\eqref{eq:PBHcriterion} for	PBH dark matter. According to Eqs.~\eqref{eq:modulation} and \eqref{eq:response}, for the power spectrum at a short-wavelength scale $\kShort$ to be significantly enhanced, we require a large long-wavelength perturbation $\curv_{\Long}$ and a large correlation $\fNL$. 

		In the lower panel of Fig.~\ref{fig:GBPowerSpecAndfNL}, we plot $\fNL(\Squeezedks)$ as a function of the long-wavelength mode $\kLong$ for two different values of the short-wavelength mode $\kShort$. The red upper curve shows $\fNL$ for a short-wavelength mode which exits the horizon after the end of USR, while the blue lower curve shows $\fNL$ for a short-wavelength mode which exits the horizon during USR. The upper and lower horizontal dashed lines show the consistency relation expectation $\fNL$ in the limit $\kLong/\kShort \rightarrow 0$.

		The numerically computed bispectrum for a short-wavelength mode which exits the horizon after USR, the red upper curve, agrees with the consistency relation. In other words, the short-wavelength perturbation $\curv_{\Short}$ retains no memory that, while it was inside the horizon, the long-wavelength perturbation $\curv_{\Long}$ outside the horizon grew in USR. This is because $\fNL(\Squeezedks)$ is set when $\kShort$ exits the horizon and the modes $\curv_{\Long}$ are already frozen at this time. Because the transformation of the bispectrum to \fncb{}   involves a subtraction of the consistency relation component, Eq.~\eqref{eq:Bzetatrans}, we conclude that short-wavelength modes which exit after the USR phase show no response to long-wavelength modes in local coordinates and therefore no enhancement of local PBH abundance.

		For a short-wavelength mode which exits the horizon during USR, the blue curve, the above logic does not hold. The numerically computed bispectrum $\fNL(\Squeezedks)$ does not agree with the consistency relation when $\curv_{\Long}$ grows outside the horizon. For such triangles, $12/5 \times \fNL \simeq 0.13$ while the consistency relation predicts $12/5 \times \fNL \simeq 0.065$.  
		
		Conversely, for the frozen $\curv_\Long$ modes that correspond to modes that exited the horizon well before USR, the consistency relation for $\fNL$ does hold. This is a successful test of our numerical computation, since in this limit the long-wavelength mode remains constant outside the horizon and just shifts the local coordinates for the small-wavelength mode along the background trajectory.

		Fig.~\ref{fig:GBPowerSpecAndfNL} also shows that when $k_\Short$ exits the horizon during USR, the near-zero of $\Delta^2_\curv$ induces a feature on $\fNL(\Squeezedks)$. This is due to the division by the power spectrum in the definition of $\fNL$, Eq.~\eqref{eq:fNLdefinition}. In particular, when $k_\Short$ exits the horizon $\curv_\Long$ has not yet reached its final (tiny) value set at the end of USR and thus a non-zero bispectrum $B_\curv(\Squeezedks)$ can be obtained. After the end of USR, $\curv_\Long$ is very small and thus $\fNL$ is amplified. The physical effect of this feature is negligible since, to obtain the power spectrum response, $\fNL$ should be multiplied by $\curv_\Long$, which has a minimum at this feature.

		More generally, the USR phase does enhance $\fNL$ relative to the consistency relation value. Hence the non-Gaussianity in \fncb{}  , $\bar{f}_{\rm NL}$, is non-zero. There is therefore an enhancement of the PBH abundances due to squeezed non-Gaussianity, which is not the case in single-field inflation on the attractor. However, for the $k_\Short$ shown in Fig.~\ref{fig:GBPowerSpecAndfNL} in blue, both $\fNL$  and $d \ln \Delta_\curv^2 / d \ln k$ are so small that, once multiplied by $\curv_\Long \ll 1$, the position-dependent effect on $\curv_\Short$ is insignificant.

		Quantitatively, we can summarize the PBH abundance enhancement in this model by choosing $\kShort$ and $\kLong$ in the USR phase, where $\fNL$ is nearly constant.  In particular, to eliminate the tilt-dependent coordinate effects on the abundance and to maximize the Gaussian part of the power spectrum, we can choose $\kShort=\kLong=\kPeak$. This triangle is not squeezed but since in USR $\fNL$ is the same for all triangle shapes, this triangle does serve as a summary statistic for local non-Gaussianity in inflection-point inflation.

		Doing so, we compute numerically that $12/5 \times \fNL(\kPeak,\kPeak,\kPeak) = 0.13$.  To obtain the response, we set $\curv_{\Long}$ to the peak value $\curv^{\rm RMS}(\kPeak) = \sqrt{\Delta^2_\curv (\kPeak)} \simeq 0.05$. Squeezed non-Gaussianity can therefore enhance the local power spectrum by a factor of at most $\sim 0.006$ and so in  inflection-point inflation, its ability 
		to enhance the local power spectrum  is negligible,
		\begin{equation}
		\frac{\Delta P_\curv}{P_\curv} \ll 1.
		\end{equation}

		In this model, we do not recover the USR squeezed limit result $12/5 \times \fNL \simeq 6$ and therefore do not enhance the small-scale power spectrum by an order-unity quantity. This is a reflection of the analytic result of Ref.~\cite{Cai:2017bxr} that transitions from a USR phase to a SR phase which are monotonically decreasing in the field velocity suppress the USR non-Gaussianity, and similar results were found numerically in Ref.~\cite{Atal:2018neu}.

		In the following sections, we will show that this suppression of non-Gaussianity is generic to transition models, except for the special case where the transition is both fast and large.

	\subsection{Infinitely Fast Transitions}
		\label{subsec:infinite}

		Infinitely fast transitions from USR to SR were considered in Refs.~\cite{Cai:2017bxr} and \cite{Pattison:2018bct},	and Ref.~\cite{Cai:2017bxr} established analytically that the final level of non-Gaussianity is sensitive to the way USR is exited.

		Specifically Ref.~\cite{Cai:2017bxr} considered the case where a pure ultra-slow-roll potential is joined to a slow-roll potential $V_{\rm SR}$ at a field position which we label $\phi_2$ for ease of generalization later:
		\begin{equation}
		\label{eq:inffast}
		V(\phi) = 
		\begin{cases}
		V_{\rm SR}(\phi_2) , &\phi > \phi_2 \quad {\rm (USR)} \\ 
		V_{\rm SR}(\phi) , &  \phi \le \phi_2 \quad {\rm (SR)}
		\end{cases}
		\end{equation}
		and hence the potential has an infinitely sharp discontinuity in slope at $\phi_2$. We call this an infinitely fast transition from USR to SR because the inflaton rolls over this discontinuity instantaneously.  $V_{\rm SR}(\phi)$ can be characterized in general by the potential slow-roll parameters for $\phi \le \phi_2$
		\begin{equation}
		\epsilon_V \equiv 
		\frac{1}{2} \left( \frac{1}{V}  \frac{d V}{d\phi}\right)^2, 
		\quad \eta_V\equiv \frac{1}{V} \frac{d^2V}{d\phi^2}.
		\end{equation}
		The transition can be characterized by the {strictly positive} amplitude parameter\footnote{The $h$ defined in \eqref{eq:h} is equivalent to Ref.~\cite{Cai:2017bxr}'s $-h/6$.}
		\begin{equation}
		\label{eq:h}
		h \equiv \sqrt{\frac{\epsilon_V (\phi_2^-)}{\epsilon(\phi_2^+)}},
		\end{equation}
		where
		\begin{equation}\label{eq:phidefNarrow}\phi_2^{\pm} \equiv \lim_{\omega\rightarrow0} \phi_2 \pm \omega,\end{equation}
		which is a ratio between the potential slow-roll parameter at the beginning of the SR phase and the Hubble slow-roll parameter $\epsilon$ at the end of the USR phase. 

		If $h=1$, the kinetic energy at the end of USR is just enough to keep the field on the attractor of the SR phase.
		The $h \ll 1$ limit therefore corresponds to the small transition, a monotonic transition from USR to the SR attractor 
		where the inflaton continues to slow down before hitting the attractor and hence the power spectrum
		continues to evolve.  
		Conversely for $h\gg 1$, Ref.~\cite{Cai:2017bxr} showed that the perturbations freeze out at $N(\phi_2)$.
		We call this a {\bf large} transition because the inflaton instantly goes from having too much kinetic energy for the potential it evolves on to suddenly having insufficient kinetic energy for a now much steeper potential
		\begin{equation}
		\label{eq:CaiCriterion} h\gg 1 \implies
		\epsilon_V
		\gg  \epsilon = \frac{1}{2} \phi'^2 \ , 
		\end{equation}		
		
		Since perturbations do not freezeout immediately for a finite value of $h$ the final level of non-Gaussianity is  not given by Eq.~\eqref{eq:USRfNL} but rather can be shown analytically to be \cite{Cai:2017bxr}
		\begin{equation} 
		\label{eq:CaifNL}
		\lim_{\kLong/\kShort\rightarrow0} \frac{12}{5} \fNL(\Squeezedks) = 2 \frac{h ( 3 h + \eta_V)}{(h+1)^2},
		\end{equation}
		for scales $\kLong$, $\kShort$ which cross the horizon during USR. 	Eq.~\eqref{eq:CaifNL} yields the USR result Eq.~\eqref{eq:USRfNL} only in the limit $h\gg1$, and thus for infinitely fast transitions the USR non-Gaussianity is conserved only when the transition is large.
		
		The enhancement of the local power spectrum is suppressed for small transitions, but we shall next see
		that it is also suppressed if the transition is not sufficiently fast.   
		Therefore, the transition needs to be large and fast to recover the USR non-Gaussianity.  
		In contrast, the inflection model of the previous
		section is an example where the transition is both small and slow.

	\subsection{Fast/Slow--Large/Small Transitions}
		\label{subsec:flat}

		In order to study in more detail the phenomenology of transient USR inflationary phases beyond the slow-small transition of \S\ref{subsec:inflection} and the infinitely fast limit of \S\ref{subsec:infinite}, we construct a toy inflationary model which begins in SR, enters a USR phase, and then transitions back to SR. We implement this with a potential 
		where the slope of an otherwise linear potential makes two transitions across  adjustable
		widths in field space
		\begin{align*}
			\frac{d V}{d\phi}(\phi) &= \frac{\beta}{2} \left[1 
		 + \tanh \left(\frac{\phi 
		 - \phi_1}{\delta_1}\right) \right]\\
		  &\quad+ \frac{\gamma}{2} \left[ 1 
		 +  \tanh \left( \frac{\phi_2-\phi}{\delta_2}\right)\right]	,\numberthis
		\end{align*}
		and hence
		\begin{align*} \label{eq:tanhpotential}
			V(\phi) &= V_0 + \frac{\beta}{2} \left[\phi 
		 + \delta_1 \log \left\{ \cosh \left(\frac{\phi 
		 - \phi_1}{\delta_1}\right)\right\} \right]\\
		  &\quad+ \frac{\gamma}{2} \left[ \phi 
		 - \delta_2 \log \left\{ \cosh \left( \frac{\phi_2-\phi}{\delta_2}\right)\right\}\right]. \numberthis
		\end{align*}
		This potential describes three phases with finite transitions, which is a natural generalization of the model~\eqref{eq:inffast} with two phases with instant transition considered in \S\ref{subsec:infinite}.
		The model parameters $\left\{\delta_1,\ \delta_2\right\}$ and $\left\{\phi_1,\ \phi_2\right\}$ determine the widths and positions of two transitions, respectively.
		The limit $\delta_1,\delta_2\to 0$ amounts to instant transitions, where the potential is composed of
		a flat plateau of amplitude $V_0$ for $\phi_2<\phi<\phi_1$ in between two linear pieces of slope $\left\{\beta,\ \gamma\right\}$, which we set positive. 
		By modifying these parameters, we can set the duration of the USR phase as well as the circumstances of its beginning and end. 
		By constructing the transition in $dV/d\phi(\phi)$ rather than in $V(\phi)$ directly, 
		we trivially obtain	a monotonic potential where the field always rolls downhill.

		Since this is a toy model that we introduce to illustrate the fast/slow  and large/small distinction, we do
		not attempt to accurately fit measurements at CMB scales or to appropriately end inflation. Once all modes which we are interested in have frozen out in the latter slow-roll phase, we end inflation by hand. By adjusting the potential before and after the plateau, one could turn this toy model into a model which can fit CMB observations and produce PBHs while ending inflation gracefully without changing the conclusions we draw below.

		For $\phi>\phi_1$ the potential has a positive slope $\simeq \beta$ and the inflaton follows the slow-roll attractor.
		In order for the inflaton to leave the slow-roll attractor and enter a USR phase, the transition must be sufficiently sharp that the inflaton enters the flat region of the potential with excess kinetic energy. Thus we fix the entry parameters $\{\beta,\ \phi_1,\ \delta_1\} = \{10^{-14},\ 0,\ 10^{-2}\}$ to guarantee such a transition. By having inflation start on the slow-roll attractor, we are freed from having to specify initial conditions during USR.
		
		The region $\phi_2<\phi<\phi_1$ marks the USR phase where the potential is approximately flat.
		We fix the amplitude of the potential in the flat plateau $V_0= 2 \times 10^{-14}$, chosen to ensure that USR modes in our fiducial model are still perturbative, i.e.~$\Delta^2_\zeta \lesssim 1$ for the durations of USR we consider here.

		Finally, for $\phi<\phi_2$ the potential has a positive tilt $\simeq \gamma$ and the inflaton returns to the slow-roll attractor.
		Among the remaining parameters $\{\phi_2,\ \delta_2,\ \gamma\}$, $\phi_2$ determines the duration of the USR period, and $\{\delta_2,\gamma\}$ set the circumstances of the exit from USR. $\phi_2$ in particular must be very finely tuned to allow several $e$-folds of USR inflation while still reaching the transition point in a reasonable amount of time. The instant transition of \S\ref{subsec:infinite} corresponds to taking $\delta_2\to0$ and to focusing on the inflaton behavior around $\phi = \phi_2$.	

		We generalize Ref.~\cite{Cai:2017bxr}'s analysis to transitions of finite width between the flat and slow-roll potentials by allowing $\delta_2 \neq 0$. We start by generalizing  the definition for the start and end of the transition, Eq.~\eqref{eq:phidefNarrow}. We choose the end of the transition $\phi_2^-$ from the potential through
		\begin{equation}
		\phi_2^- \equiv \phi_2 - 2 \delta_2.
		\end{equation}
		The beginning of the transition, $\phi_2^+$, is not simply $\phi_2+2\delta_2$ since the USR phase persists while $\epsilon_V \ll \epsilon$. Instead, we choose to define the beginning of the transition through the deviation from the USR analytic solution, 
		\begin{equation}
		1-\frac{\phi'_{\USR}}{\phi'}\bigg\vert_{\phi_2^+}  = 0.05 \times \left(1-\frac{\phi'_{\USR}}{\phi'}\bigg\vert_{\phi_2^-}\right),
		\end{equation}
		where $\phi'_{\USR}$ is the analytic solution for the field velocity in USR, and $\phi'$ is the actual field velocity, which is evaluated numerically.
		By computing the field velocity deviation relative to the change at the end of the transition $\phi_2^-$ we guarantee that $\phi_2^+$ can be defined even for small and fast transitions.
				
		In other words, $\phi_2^-$ is roughly where the potential completes its transition, and $\phi_2^+$ is roughly where the field velocity begins to leave the USR solution. While the specific criteria chosen here are arbitrary, they are useful for classifying transition regimes and in the $\delta_2 \rightarrow 0$ limit the choices here return the limit Eq.~\eqref{eq:phidefNarrow} up to percent-level factors. 

		From these definitions for $\phi_2^+$ and $\phi_2^-$ we compute $h$ by evaluation of Eq.~\eqref{eq:h} and we quantify the duration of the transition from USR inflation to the beginning of the relaxation process
		\begin{equation}
		\label{eq:dN}
		d_N \equiv N(\phi_2^-) - N(\phi_2^+).
		\end{equation}

		\begin{figure}[t]
		\psfig{file=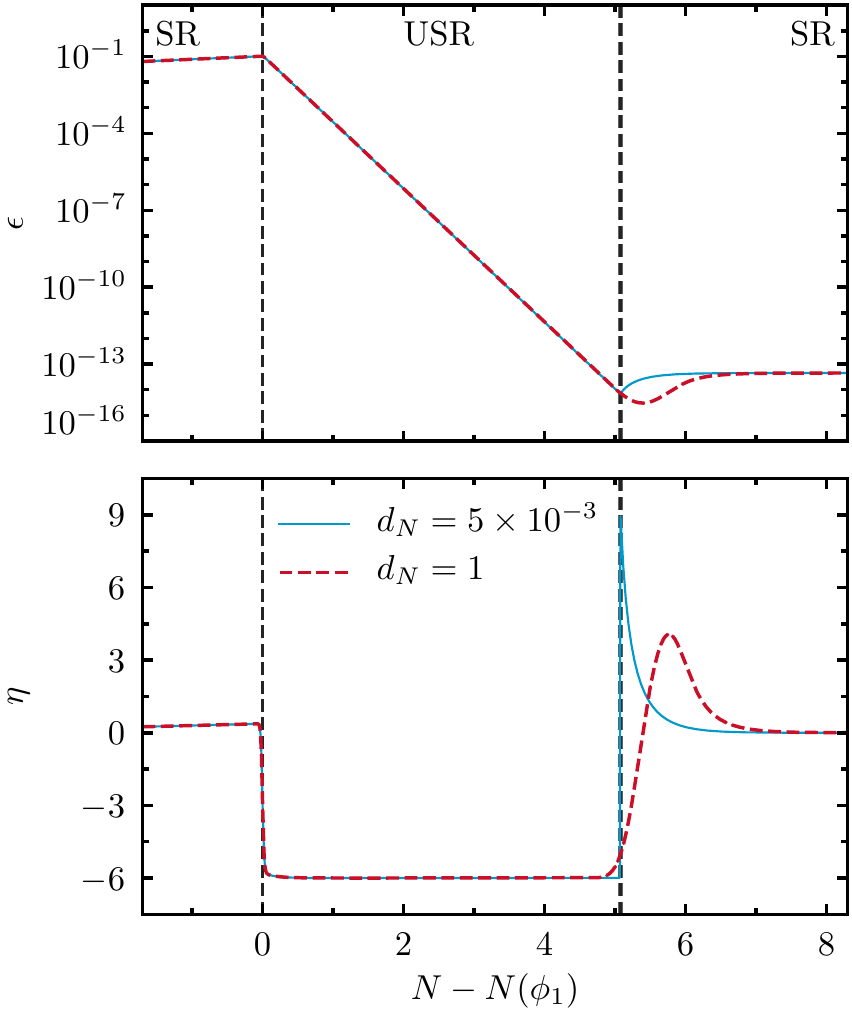}
		\caption{Transition model background parameters $\e \equiv - d\ln H/d N$ and $\eta \equiv d\ln\e / d N$ for a fast vs.~slow transition, denoted by the $e$-fold width $d_N$, with a fixed large transition ($h=2.5$). The vertical lines mark $N(\phi_1)$ and $N(\phi_2^+)$.  The specific parameter choices used for these models are described in \S\ref{subsec:flat}. 
		}
		\label{fig:USRBackground}
		\end{figure}

		The situation of \S\ref{subsec:infinite} corresponds to the limit $d_N \rightarrow 0$ and we now generalize this result by exploring the impact of the duration $d_N$ on the resultant non-Gaussianity. 
		In Fig.~\ref{fig:USRBackground}, we show the background parameters $\e$ and $\eta$ for two models with a large transition $h$ = 2.5, one fast ($d_N = 5\times 10^{-3}$) and the other slow  ($d_N = 1$). For these models $\gamma$ is fixed at $6 \times 10^{-21}$, while the fast transition has $\left\{\phi_2 ,\ \delta_2\right\} = \left\{-0.1580281699,\ 2.12 \times 10^{-10} \right\}$ and the slow transition has $\left\{\phi_2 ,\ \delta_2\right\} = \left\{-0.1580282187,\ 3.6\times10^{-8} \right\}$.  Notice the amount of fine-tuning in $\phi_2$ required to achieve subpercent-level control of the transition amplitude and duration.

		Defining $n \equiv N - N(\phi_2^-)$ as the positive increasing number of $e$-folds elapsed since the potential transition, in the fast transition limit $d_N\rightarrow0$ the solution for $\eta$ after the transition point behaves according to the analytic result \cite{Cai:2017bxr} 
				\begin{equation}
		\label{eq:AnalyticEta}
		\eta(n > 0)=\frac{6 (h-1)}{1+h(e^{3 n}-1)}~.
		\end{equation}
		This is reflected in the behavior of the blue solid curve in Fig.~\ref{fig:USRBackground}, which behaves as Eq.~\eqref{eq:AnalyticEta} up to a ${\sim0.3\%}$ difference in the $h$ parameter as defined here compared to the $h$ parameter in the exact $\delta_2 \rightarrow 0$ limit. 
		
		On the other hand, in the case where $d_N$ is large, the red dashed curve of Fig.~\ref{fig:USRBackground}, the numerical solution for $\eta$ deviates significantly from this analytic form. This can be understood by Taylor expanding Eq.~\eqref{eq:AnalyticEta} around the transition point $n=0$,
		\begin{equation}
		\eta (n > 0) =  6(h-1) \left( 1 - 3 h n \right)   + \O\left(n^2\right),
		\end{equation}
		from which we can see that after $\phi_2^-$, $\eta$ evolves on a timescale $n \sim  h^{-1}$. Thus if the transition timescale $d_N$ is larger than this timescale, the evolution of $\eta$ will differ from the analytic solution \eqref{eq:AnalyticEta}.

		The behavior of $\eta$ is important because it comes directly into the source of squeezed non-Gaussianity in the in-in formalism, Eq.~\eqref{eq:cubicLagrangianSqueezed}, and controls the freezeout of perturbations through  the evolution of $\epsilon$ in Eq.~\eqref{eqn:Rprimesoln}. Therefore, the timescale $d_N$ plays an important role in 
		changing the  non-Gaussianity produced in USR.

		\begin{figure}[t]
		\psfig{file=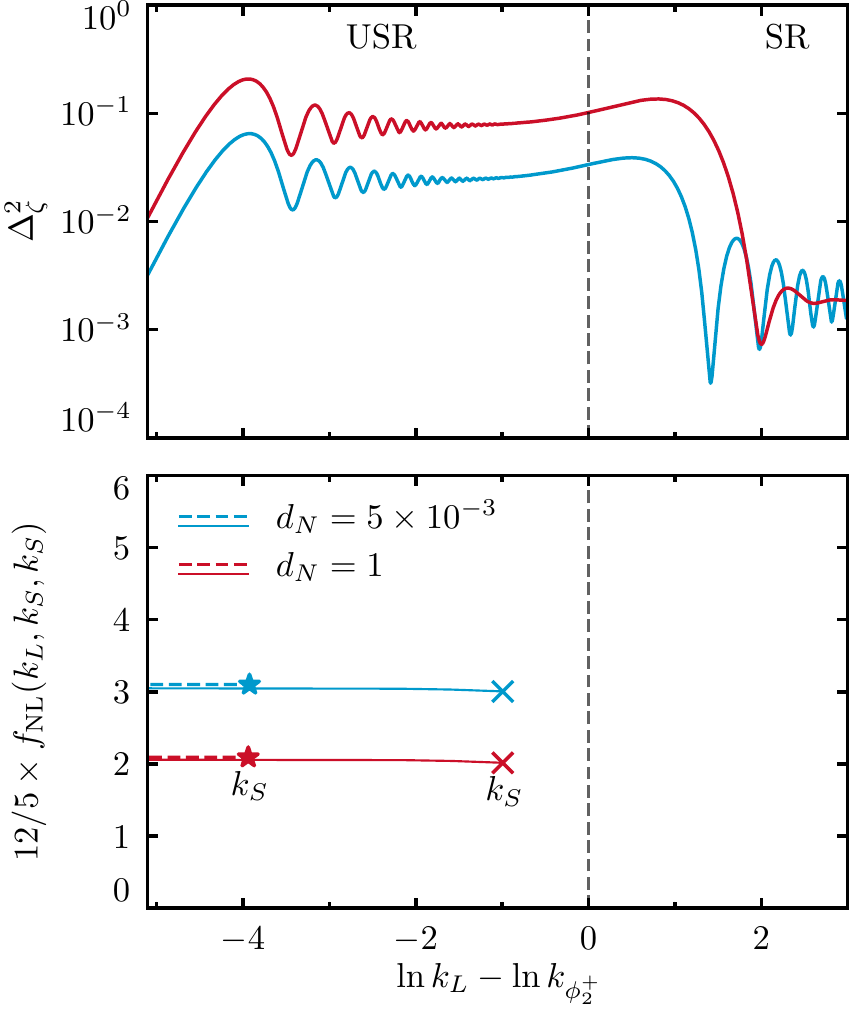}
		\caption{Transition model power spectrum $\Delta^2_\curv$ and non-Gaussianity parameter $\fNL(\Squeezedks)$ for the large-fast (blue) and large-slow (red) transition models of Fig.~\ref{fig:USRBackground}.   The left edge 
		corresponds roughly to a mode which exits the horizon at the beginning of USR. Conventions for displaying $\fNL$ are the same as in Fig.~\ref{fig:GBPowerSpecAndfNL} except both  $\kShort$ values for each model cross the horizon before the end of USR (vertical line, $k_{\phi_2^+}$) (see \S\ref{subsec:flat} for further discussion).}
		\label{fig:USRPowerSpecAndfNL}
		\end{figure}

		The upper panel of Fig.~\ref{fig:USRPowerSpecAndfNL} shows the power spectra for the same large-fast and large-slow models as Fig.~\ref{fig:USRBackground}. Once more, the red upper curve shows the slow transition, while the blue lower curve shows the fast transition. The vertical dashed line shows a mode which exits the horizon at the onset of the transition $N(\phi_2^+)$.

		The power spectra in these models show many of the same features as the inflection-point power spectrum, Fig.~\ref{fig:GBPowerSpecAndfNL}, and therefore we focus on the USR and transition regions of the plot, where unlike the inflection model these models have power spectrum plateaus for modes which exit during USR. This plateau is modulated by small oscillations sourced by the first feature in the potential. The USR to SR transition feature in the potential also induces power spectrum oscillations, with the fast transition model having more oscillations due to the sharper source. The slow transition model has a larger power spectrum than the fast transition as $\e$ reaches a lower level in this model (see Fig.~\ref{fig:USRBackground}) and thus the modes grow more.

		The lower panel of Fig.~\ref{fig:USRPowerSpecAndfNL} shows the non-Gaussianity for these models for two different values of the short-wavelength mode, one marked with a cross and the other a star and both exiting during the USR phase, as a function of the long-wavelength mode, curves with correspondingly solid and dashed lines. All triangles yield the same value of $\fNL$ when the legs exit during the USR phase up to corrections of order $\kShort/k_{\phi_2^+}$, consistent with the result in the inflection model \S\ref{subsec:inflection} and the exact USR result of \S\ref{sec:usr}. However, for neither model does the level of the non-Gaussianity agree with the analytic result for USR Eq.~\eqref{eq:USRfNL}, $5 \fNL /12= 6$. This is the result of Ref.~\cite{Cai:2017bxr}, that the residual level of $\fNL$ depends on the value of the transition parameter $h$, and in particular the fast transition model yields the result expected from Eq.~\eqref{eq:CaifNL} for a transition with $h=2.5$, $\fNL \simeq 3.1$.
		However, the slow model has the same $h=2.5$ as the fast model, yet a smaller value of $\fNL$. This is due to the slow nature of the transition in the $d_N=1$ model. Fast transitions yield Eq.~\eqref{eq:CaifNL}, while slow transitions suppress the non-Gaussianity.

		We model this effect with the ansatz that a given transition length $d_N$ sets an upper bound to the transition amplitude, independent of $h$. We can define an effective transition amplitude
		\begin{equation}
		\label{eq:ansatz}
		h_{\rm eff} \equiv \left[ \left(\frac{1.5}{d_N}\right)^{-3}+ h^{-3} \right]^{-1/3}
		\end{equation}
		where the exponent serves merely to interpolate between the two limits and the factor of $1.5$ comes from calibrating the results to the form of \eqref{eq:CaifNL}
		\begin{equation}
		 \frac{12}{5} \fNL(h_{\rm eff}) =  2 \frac{h_{\rm eff} ( 3 h_{\rm eff} + \eta_V)}{(h_{\rm eff}+1)^2},
		 \label{eq:fNLhEff}
		\end{equation}
		where in this toy model with a linear SR potential we set $\eta_V=0$.			

		\begin{figure}[t]
		\psfig{file=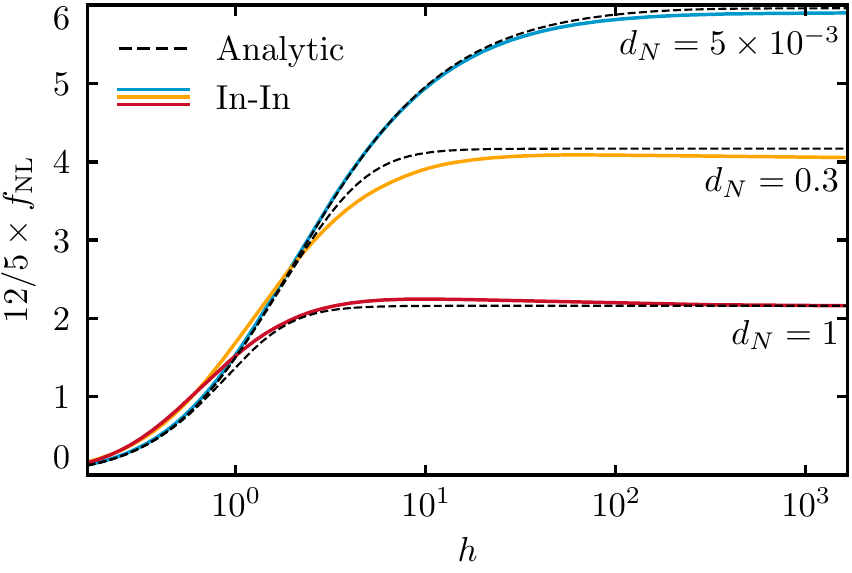}
		\caption{Transition model non-Gaussianity parameter $\fNL$ as a function of amplitude $h$ for various $e$-fold widths $d_N$ with both $k_\Long$ and $k_\Short$ exiting the horizon during USR.  Colored lines show the numerical in-in computation and dashed black lines show the calibrated analytic prediction from Eqs.~\eqref{eq:ansatz}~and~\eqref{eq:fNLhEff}.   
		To reach the USR result $12\fNL /5 = 6$, the transition must be large $h\gg 1$ and
		fast $d_N \ll 1$ (see \S\ref{subsec:flat} for further discussion). 
		}
		\label{fig:fNLvsh}
		\end{figure}

		\begin{figure*}[t]
		\psfig{file=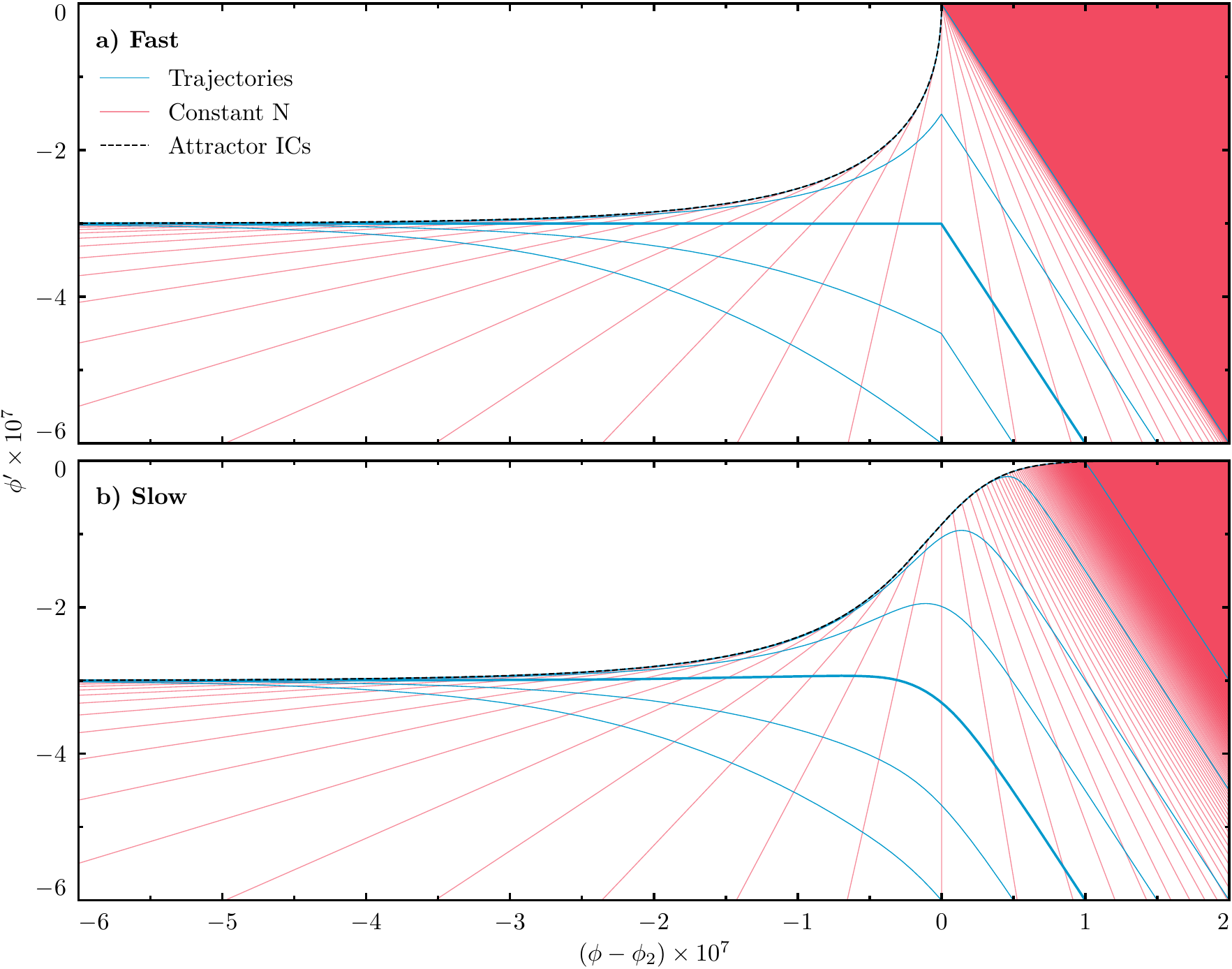}
		\caption{ Phase-space diagram for the USR to SR transition models with model trajectories (blue lines) and
		constant $e$-fold surfaces (red lines) relative to the transition at $\phi_2$. The top panel a) has $\delta_2 = 2.12\times10^{-10}$, such that most trajectories correspond to  fast transitions, while the bottom panel b) has $\delta_2 = 3.6\times10^{-8}$,
		such that most trajectories correspond to  slow transitions.   In each panel, higher trajectories represent larger transitions, with  $h=1$ as the large-small dividing line (thick blue).  
			 Constant $N$ surfaces become space-filling in the top right corners. 
			 A large $\fNL$ requires a fast-large transition as can be visualized by $\delta N$, the change
			 in $e$-folds given a shift in the initial field $\delta \phi_i$ that takes the local background to
			 a new trajectory (see \S\ref{subsec:flat} for further discussion).}\label{fig:phasespaceTransition}
		\end{figure*}

		\begin{figure}[t]
		\psfig{file=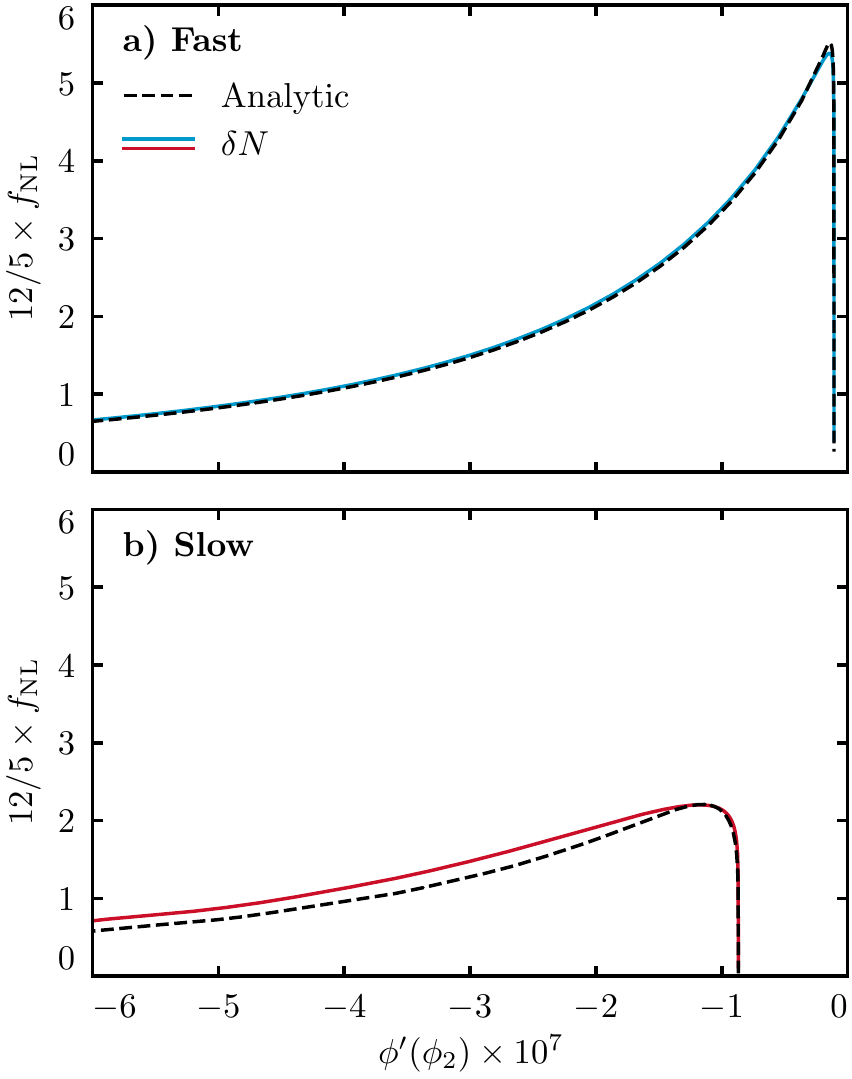}
		\caption{Non-Gaussianity for trajectories in the corresponding phase spaces of Fig.~\ref{fig:phasespaceTransition} computed via the $\delta N$ formula~\eqref{eq:deltaN} compared to our calibrated analytic result Eqs.~\eqref{eq:ansatz} and~\eqref{eq:fNLhEff}. The horizontal axis here corresponds to the vertical axis of Fig.~\ref{fig:phasespaceTransition}. Only for fast and large transitions does the transition model reproduce the USR result, and for fixed $\delta_2$ there is a maximum value of $\fNL$ attainable. 	
		}
		\label{fig:deltaN}
		\end{figure}

		In Fig.~\ref{fig:fNLvsh}, we compare the ansatz \eqref{eq:ansatz} coupled with the analytic formula  \eqref{eq:fNLhEff} (black dashed curves) to the full numerical in-in computation of the bispectrum (colored solid curves) for modes that exit the horizon during USR
		as a function of the transition amplitude $h$ for three different transition speeds $d_N$.  Every point along these lines corresponds to a different set of parameters for our toy model. 

		The analytic formula \eqref{eq:ansatz} agrees well with the in-in computation for all values of $h$ and $d_N$. For small-slow transitions (small $h$ and large $d_N$), where $\fNL$ is small and becomes proportional to the potential slow-roll parameters on the attractor, there is a slight difference between the numerical and analytic $\fNL$. This we attribute to small differences in the value of $h$ as defined for the infinitely fast transitions and as defined for slow transitions since the errors  decrease for smaller values of $d_N$.  For large transitions $h\gg1$, where $\fNL$ is largest, the analytic ansatz produces a slight overestimate of $\fNL$ as the transitions become faster.  This is due to non-linearities between the true $h_{\rm eff}$ and $1/d_N$ in the large $h$ limit which our ansatz does not model.
		Since these differences are minor,
		we conclude from the analytic formula that to produce a large level of non-Gaussianity after the transition from USR to SR requires $h_{\rm eff} \gg 1$ and thus the transition must be large, ${h \gg 1 }$, and fast, ${d_N \ll 1}$.

		Just as in the SR (\S\ref{sec:nogo}) and exact USR case (\S\ref{sec:usr}), the effect on $\fNL$ of the transition from USR to SR can be understood visually from the way phase-space trajectories intersect constant $N$
		surfaces. Fig.~\ref{fig:phasespaceTransition}\hyperref[fig:phasespaceTransition]{a} shows the phase-space 
		trajectories (blue lines) and constant $N$ surfaces (red lines) for a narrow $\delta_2 = 2.12 \times 10^{-10}$ and therefore faster transitions, such that the fast model of Fig.~\ref{fig:USRBackground} and Fig.~\ref{fig:USRPowerSpecAndfNL} corresponds to a trajectory in this space. Fig.~\ref{fig:phasespaceTransition}\hyperref[fig:phasespaceTransition]{b} shows the phase space for a wider $\delta_2 = 3.6 \times 10^{-8}$ and therefore slower transitions, and the slow model of Fig.~\ref{fig:USRBackground} and Fig.~\ref{fig:USRPowerSpecAndfNL} evolves through this space.

		Trajectories near the top of each panel have the inflaton speed up after the transition and hence
		have a large $h$, with the large-small dividing line of $h=1$ denoted by thick blue lines.
		Notice also that the
		union of the panels of Fig.~\ref{fig:phasespaceSR+USR} gives the limit of infinitely fast transitions,
		with the exception that here constant $N$ surfaces are plotted relative to the transition feature
		$N(\phi_2)$ rather than the end of inflation $N=0$.  Trajectories are evenly spaced in $\phi$ at the point where they cross the bottom edge at $\phi' =-6$ in a range that reflects a reasonable amount of USR $e$-folds as we describe next.

		Due to the smooth nature of the potential \eqref{eq:tanhpotential}, for any finite $\phi>\phi_2$ the potential slope $dV/d\phi$ has a finite positive value. Thus, unlike in the exact USR case (Fig.~\ref{fig:phasespaceSR+USR}\hyperref[fig:phasespaceSR+USR]{b}) or the infinitely fast case, all trajectories 
		with any finite $\phi'$ for $\phi>\phi_2$ will eventually cross $\phi_2$.   

		Black dashed curves in Fig.~\ref{fig:phasespaceTransition} depict the envelope of such trajectories, neglecting stochastic effects, and correspond to initial conditions on the attractor on the very nearly flat potential.  Consequently, the constant $e$-fold surfaces become increasingly tightly packed and eventually space-filling, in contrast to the empty upper right triangle in Fig.~\ref{fig:phasespaceSR+USR}\hyperref[fig:phasespaceSR+USR]{b}.  We choose not to continue showing trajectories which take such large numbers of $e$-folds to traverse the nearly flat plateau of the potential.

		By the same $\delta N$ arguments of \S\ref{sec:usr} we can immediately see from these phase spaces why a large ($h \gg 1$) transition is necessary to conserve the USR non-Gaussianity. Here $\delta N$ refers to the change in the total number
		of $e$-folds elapsed to a fixed field position on the SR side for a shift in the initial field
		 position $\delta \phi_i$ on the USR side which then shifts the whole trajectory. Note that
		 $\delta N$ combines the change from the USR and SR sides.
		 
		Let us first consider the fast case in the top panel. Around a central trajectory with large $h$ (upper trajectories), 
		the crossing rate $\partial N/\partial\phi_i$  is strongly asymmetric to the sign of $\delta\phi$,
		i.e.~there is a large second derivative $\partial^2 N/\partial \phi_i^2$
		and hence a large $f_{\rm NL}$ according to
		 Eq.~\eqref{eq:deltaN}. This is due to the much larger contribution to the rate of surfaces crossed in
		 the USR side where the asymmetry is larger than the SR side where the asymmetry is small. 

		 On the other hand, the asymmetry around a small $h$ trajectory (lower trajectories) is small and therefore the non-Gaussianity is small. This is due to the smaller  contribution to $\partial N/\partial \phi_i$ on the USR side relative to the SR side.  In other words the power spectrum continues to grow on the
		SR side at small $h$, which suppresses the non-Gaussianity from the USR side.
		
		By comparing the fast and slow cases, we can visually see that the transition duration sets an effective maximum transition amplitude $h_{\rm eff}$. Above a certain value of $h$, the trajectory joins the slow-roll attractor before the transition and therefore will have a highly suppressed non-Gaussianity comparable to the small-slow transition of \S\ref{subsec:inflection}.

		In Fig.~\ref{fig:deltaN} we formalize these heuristic  arguments by computing the $\delta N$ formula~\eqref{eq:deltaN} for different trajectories in these phase spaces. We organize the trajectories by their velocity at $\phi_2$, and thus the horizontal axis of Fig.~\ref{fig:deltaN} corresponds to the vertical axis of the Fig.~\ref{fig:phasespaceTransition} at the transition point. We then also compare the $\delta N$ result to our analytic expectation for $\fNL$, Eqs.~\eqref{eq:ansatz} and \eqref{eq:fNLhEff}.

		The $\delta N$ computation cross-validates our analytic formula which was calibrated to the in-in
		calculations, showing excellent agreement for all methods of computation across the fast-slow and 
		large-small transition space.  The $\delta N$ computation thus also confirms that a large $\fNL$ requires a large-fast transition. 
		
		For large $\phi'(\phi_2)$, after the peak non-Gaussianity, there is a sharp cliff in Fig.~\ref{fig:deltaN} beyond which $\fNL$ becomes suppressed. This cliff corresponds to trajectories which reach the attractor before the transition, thus inhabiting the space-filling regions of Fig.~\ref{fig:phasespaceTransition}, and the edge of the
		cliff is given by the  black dashed envelope of Fig.~\ref{fig:phasespaceTransition}. These trajectories are slow transitions even though they have the same narrow field space width $\delta_2$ and appear in the `fast' panel. These
		cases behave in the same way as those in the `slow' panel once the duration of the transition in $e$-folds $d_N$ is accounted for. 

		Of course even for a large and fast transition, for which $5/12 \ \fNL \rightarrow 6$, the response of the small scale power spectrum to the long-wavelength mode is still dependent on the value of the long-wavelength mode, and in particular the argument of \S\ref{sec:usr} still holds, that 
		\begin{equation}
		\frac{\Delta P_\curv}{P_\curv} \lesssim 1,
		\end{equation}
		unless the model already produces PBHs with a Gaussian distribution.

\section{Conclusion}
	\label{sec:conclusion}
	Canonical slow-roll inflation cannot produce primordial black holes in a large enough quantity to be the dark matter. While perturbations do exhibit a small level of local non-Gaussianity which couples short-wavelength PBH fluctuations to the long-wavelength modes they live in and can in principle enhance local abundances at peaks of long-wavelength modes, transforming to a freely-falling coordinate system shows that locally measured PBH abundances are completely insensitive to this non-Gaussianity because it is generated by the reverse coordinate transformation to begin with. 

	Any confirmation that the dark matter is in the form of PBHs would rule out canonical slow-roll inflation. The only way to rescue canonical inflation would be to violate slow roll, and a phase of ultra-slow-roll inflation after CMB scales exit the horizon is the natural way to do this. We showed by gauge transformation and by the $\delta N$ formalism, which can be illustrated graphically and contrasted to the SR case, that local non-Gaussianities are large in the USR phase when perturbations freeze out instantly at some fixed field position. 

	The same coordinate transformation machinery as in slow roll confirms that squeezed USR non-Gaussianities can locally enhance PBH abundances. However the effect is very mild, giving at most an order unity enhancement of the local power spectrum. Such enhancements can only make models that are already on the borderline of succeeding to produce PBHs as the dark matter under Gaussian assumptions  actually succeed.
	For such cases, generally a small change in parameters that prolong the USR phase would equally well produce  PBHs under Gaussian assumptions. 

	Even more importantly, the USR phase has to end in some way. Ref.~\cite{Cai:2017bxr} established that the non-Gaussianity is very sensitive to how this period ends using cases where the transition is
	infinitely fast. By exact computation in the in-in formalism and validation with the $\delta N$ formalism, we mapped the entire range of possible endings to USR to show that only a small class of transitions conserves the large USR non-Gaussianity through the transition to slow roll. These are the transitions which are fast, in that the potential exhibits a sharp feature that is traversed by the inflaton in much less than an $e$-fold, and large, in that the inflaton needs to gain significant velocity after transiting the feature. All other types of transitions suppress the non-Gaussianity significantly.

	Note that while we have computed the full local non-Gaussianity, it is only the squeezed non-Gaussianity which we have shown has a negligible effect on whether a model produces PBH dark matter. Understanding the effect of other mode couplings requires a full knowledge of the probability distribution function of the density perturbation taking into account the contributions from all orders of non-Gaussian correlators of every shape. Work in this direction has been pursued in a variety of contexts (see, e.g., \cite{Saito:2008em,Smith:2011,Byrnes:2012yx,Ferraro:2014jba,Young:2015kda,Tada:2015noa,Franciolini:2018vbk}), but it remains a challenging problem and direction for future work.

	Nonetheless the conclusion that producing primordial black holes as dark matter in canonical single-field inflation requires a complicated and fine-tuned potential shape with a transient violation of slow roll is robust to the inclusion of local squeezed non-Gaussian effects.

\acknowledgments

	We thank Macarena Lagos, Meng-Xiang Lin, and Marco Raveri for fruitful discussions.
	SP and WH were supported by U.S.\ Dept.\ of Energy contract DE-FG02-13ER41958, NASA ATP NNX15AK22G, and the Simons Foundation.  SP was additionally supported by the Kavli Institute for Cosmological Physics at the University of Chicago through grant NSF PHY-1125897 and an endowment from the Kavli Foundation and its founder Fred Kavli.
	HM was supported by Japan Society for the Promotion of Science (JSPS) Grants-in-Aid for Scientific Research (KAKENHI) Nos.\ JP17H06359 and JP18K13565.

\appendix

\section{Bispectrum Computation}
	\subsection{Numerical Methods}
		\label{app:numerics}

		We describe in this appendix the numerical computation of  the background evolution, the modefunctions, and the cubic interactions during inflation in order to study the precise predictions of inflationary scenarios which transition between SR and USR phases.

		After numerically solving for the background evolution using the canonical Klein-Gordon and Friedmann equations, we use the quadratic action for the comoving curvature perturbation $\curv$
		\begin{equation}
		\label{eq:quadraticAction}
		S_2 = \int \difffour{x} \ \L_2,
		\end{equation}
		with the quadratic Lagrangian density 
		\begin{equation}
		\label{eq:L2} 
		\L_2 \equiv a^3 \epsilon \left[ \dot{\zeta}^2 - \frac{1}{a^2} \left(\pa \zeta\right)^2 \right] , 
		\end{equation}
		to determine the evolution of the modefunctions through the Mukhanov-Sasaki equation of motion
		\begin{equation}
		\label{eq:MukhanovSasaki}
		\frac{1}{a^2\epsilon}\frac{d}{ds} \left( a^2 \epsilon \frac{ d \curv_k}{d s} \right)  + k^2 \curv_k = 0,
		\end{equation}
			where  $s \equiv \int_{t}^{t_{\rm end}} dt / a$, with  Bunch-Davies initial conditions at $ks\gg 1$ of the form
		\begin{equation}
		\label{eq:BunchDavies}
		\curv_k^0 = \frac{1}{2 a\sqrt{ k \epsilon}} \left(1 + \frac{i}{k s}\right) e^{i k s}.
		\end{equation}

		From these modefunctions we can construct the Fourier space interaction-picture field operators
		\begin{equation}
		\hat{\curv}^I_{\bf{k}} = \curv_k \hat a({\bf k}) + \curv_k^* \hat a^\dag(-{\bf k}) ,
		\end{equation}
		where the creation and annihilation operators satisfy the usual commutation relation
		\begin{equation}
		[\hat a({\bf k}), \hat a^\dag({\bf k}')] = (2\pi)^3 \delta({\bf k}-{\bf k}').
		\end{equation}
		The power spectrum can be evaluated from the modefunctions at a time $t_*$ taken to be after all the relevant modes have frozen out (e.g., after the end of the USR phase) as 
		\begin{align}
		\langle \hat{\curv}^I_{\textbf{k}}(t_*)\hat{\curv}^I_{\textbf{k}'}(t_*)\rangle & = 
		 (2\pi)^3\delta^{3}(\textbf{k}+\textbf{k}') |\curv_k(t_*)|^2  
		\nonumber \\ &\equiv  (2\pi)^3\delta^{3}(\textbf{k}+\textbf{k}') P_\curv (k). 
		\end{align}

		The tree-level three-point correlation function is then computed in the in-in formalism as \cite{Adshead:2013zfa,Maldacena:2002vr,Weinberg:2005vy,Adshead:2009cb}
		\begin{align*}
		\label{eq:inin}
		&\langle \hat{\curv}_{\bf{k_1}} \hat{\curv}_{\bf{k_2}} \hat{\curv}_{\bf{k_3}}  \rangle  \numberthis \\
		&\simeq 2  \re{-i \int_{-\infty(1+i\epsilon)}^{t_*} \diff{t} \langle \hat{\curv}^I_{\bf{k_1}}(t_*) \hat{\curv}^I_{\bf{k_2}}(t_*) \hat{\curv}^I_{\bf{k_3}}(t_*) H_{I}(t) \rangle}, 
		\end{align*} 
		where $\hat{\curv}$ is the full field operator and in which the interaction Hamiltonian $H_I$ can be calculated at cubic order from the cubic Lagrangian $\L_3$ by $H_I \simeq - \int \diffcubed{x} \L_3$ \cite{Adshead:2008gk}. $\L_3$ itself is given by \cite{Adshead:2013zfa,Passaglia:2018afq}
		\begin{align*}
		 \label{eq:cubicLagrangianSqueezed}
		 \L_3 \equiv& \ a^3 \e 
		 \frac{\diff{}}{\diff{t}} \left(\e + \frac{\eta}{2} \right) \curv^2 \dot{\curv}
	     - \frac{\diff{}}{\diff{t}} \left[ a^3 \e \left(\e + \frac{\eta}{2} \right) \curv^2 \dot{\curv}\right] \\
	     &+ \e \curv(\Ha_2 + 2 \L_2) \\
	     &- \frac{\diff{}}{\diff{t}} \left[ \frac{a^3 \e}{H} \curv \dot{\curv}^2 + a^3 \frac{\e^2}{2 H} \dot\curv  \pa_a\curv  \pa_a(\partial^{-2} \dot\curv) \right],
		 \numberthis
		\end{align*}
		where $\Ha_2$ is the quadratic Hamiltonian density
		\begin{equation}\Ha_2 = a^3 \e \left[ \dot{\curv}^2 + \frac{1}{a^2} \left(\pa \curv\right)^2 \right].\end{equation}
		We have neglected here operators which do not contribute to the squeezed limit and for a canonical scalar (cf.~\cite{Passaglia:2018afq}).  We have also neglected here boundary operators whose contribution at $t_*$
		is  suppressed by relative factors of $k/ a H$.  Similarly, terms in the third line of \eqref{eq:cubicLagrangianSqueezed} are suppressed by the extra $\dot \curv$ factor so long as $t_*$ is taken to be  after all modes have frozen out, i.e. after the transition, in the context of transient USR.   We take this approach in the main paper.  In  App.~\ref{app:gauge}, we consider this extra boundary contribution if the correlator is evaluated during USR.

		From translational and rotational invariance the three-point correlator is related to the bispectrum $B_\curv$ through
		\begin{equation}
		\langle \hat{\curv}_{\bf{k_1}} \hat{\curv}_{\bf{k_2}} \hat{\curv}_{\bf{k_3}}\rangle
		= (2\pi)^3 \delta^{3}({\bf{k_1}}+{\bf{k_2}}+{\bf{k_3}}) B_{\curv}(k_1,k_2,k_3), 
		\end{equation}
		from which is constructed the conventional non-Gaussianity parameter, Eq.~\eqref{eq:fNLdefinition}.

		In this work we solve these formulae numerically to compute the non-Gaussian response of perturbations in generic inflationary scenarios. We use these numerical results to calibrate an analytic formula for models with smooth USR to SR transitions in  \S\ref{subsec:flat}.  In addition, we exploit the $\delta N$ formalism, which requires only numerical solutions for various background evolutions,  to validate and visualize our results. In the circumstance of pure USR inflation an analytic approach can be followed which we consider next.

	\subsection{USR Gauge vs.~Field Redefinition}
		\label{app:gauge}

		Under some conditions the bispectrum can be calculated much more simply than the numerical approach presented in App.~\ref{app:numerics}. In this appendix, we compute the bispectrum in USR (\S\ref{sec:usr}) analytically. This can be done 
		by transforming between spatially flat and comoving gauges since the flat potential implies negligible field interactions in the former. We explicitly show this to be the case below and clarify subtleties due to the shift in evaluation time induced by the gauge
		transformation on cubic interactions from the boundary terms.   This gauge transformation is sometimes
		phrased as a field redefinition in the literature \cite{Maldacena:2002vr},
		 since the dependence on the
		evaluation time drops out in ordinary slow roll where the curvature is frozen outside the horizon. 
		Even in USR the three-point correlations can still 
		can be deduced directly from the two-point correlations from this perspective (see., e.g., Refs.~\cite{Namjoo:2012aa,Martin:2012pe,Chen:2013aj,Chen:2013eea}) but our approach clarifies the role of the cubic interaction terms involved in their direct computation.

		We start the computation in spatially flat-gauge, where the bispectrum computation is trivial, and then transform to comoving gauge, which is the relevant gauge for observations. The relationship between the time coordinate in these gauges is 
		\begin{equation}
		\label{eq:timetransform}
		t_{\SF} = t_{\U} + \frac{\curv}{H}
		\end{equation}
		where $t_{\text{SF}}$ denotes the time coordinate in spatially flat gauge, $t_{\text{U}}$ the time coordinate in comoving gauge, and $\curv$ is the comoving-gauge curvature perturbation.\footnote{In USR, $\curv$ grows and so does 
		the difference between the time coordinate  of spatially flat gauge and comoving gauge.  Hence
		even though a growing curvature violates the separate universe criteria for comoving slicing \cite{Hu:2016wfa}, spatially flat slicing observers are nearly freely-falling and see superhorizon field perturbations as a local FLRW  background.} 	
		 Letting $\curv_N$ be the  rescaled field perturbation in spatially-flat gauge, $\curv_N \equiv -H \delta \phi / \dot{\phi}$, the relationship between the perturbations is
		\begin{equation}
		\label{eq:fieldtransform}
		\curv = \curv_{N} + f(\curv_{N})
		\end{equation}
		where
		\begin{equation}
		f(\curv_{N}) = \frac{\eta}{4}\curv_N^2+\frac{1}{H}\curv_N\dot{\curv}_N
		\label{eq:nonlinearterm}
		\end{equation}
		to the lowest order in $\curv_{N}$ and up to terms suppressed outside the horizon by factors of $k/(a H)$. Eq.~\eqref{eq:fieldtransform} can be viewed equivalently as simply a field redefinition.

		In spatially flat gauge, all interactions should be suppressed by $\e$, which is extremely small in USR. We therefore assume for now that the spatially-flat gauge fields are free fields,
		\begin{equation}
		\langle \curv_N \curv_N \curv_N \rangle = 0.
		\end{equation}
		We can use Eq.~\eqref{eq:fieldtransform} to find that in USR
		\begin{align*}
		\label{eq:analyticresult}
		\langle \curv_{k_1} \curv_{k_2}  \curv_{k_3}\rangle &= (2 \pi)^3 \delta^3(k_1 + k_2 + k_3)\\ &\quad \times  \frac{3 H^4 (k_1^3 + k_2^3 + k_3^3)}{16 \e_{\text{end}}^2 k_1^3 k_2^3 k_3^3},
		\numberthis
		\end{align*}
		where we have evaluated correlators of the form $\langle \curv \curv \rangle$ and $\langle \curv \dot{\curv} \rangle$ using	\eqref{eq:BunchDavies} with the USR scalings $\e=\e_* (a_*/a)^6$ and  $s \simeq (a H)^{-1}$  which yield (e.g. \cite{Namjoo:2012aa})
		\begin{equation}
		\label{eq:USRmodefunction}
		\curv_k = -\frac{i H}{ 2 \sqrt{\e_* k^3} }\left( \frac{a}{a_*}\right)^3 
		\left(1 - i k s\right) e^{i k s},
		\end{equation} 
		and the subscript $*$ here indicates the values of the parameters at an arbitrary reference time during the USR phase. Note that choosing $a_*=a_{\rm end}$, the end of the USR phase, leads to Eq.~\eqref{eq:Delta2USR} for the power spectrum as well.
	
		This result for the bispectrum leads to the well-known result  that the squeezed bispectrum in USR inflation violates the consistency relation prediction of zero. This result is often computed from a field redefinition of the form of Eq.~\eqref{eq:fieldtransform}. However, this can be confusing as the basis 
		of this derivation, $\langle \curv_N \curv_N \curv_N \rangle=0$, is a statement about perturbations of a free field in spatially flat gauge rather than a statement about interactions of the redefined field  in comoving gauge. To see this, we can take opposite approach and start with the action in comoving gauge, transform it to spatially flat gauge,
		and show how and why $\langle \curv_N \curv_N \curv_N \rangle = 0$.

		Neglecting all terms suppressed by factors of $\e$, factors of $k/(a H)$, or irrelevant in the squeezed limit, the action for the comoving curvature perturbation $\zeta$ in USR is given by
		\begin{align*}
		\label{eq:S23c}
		S[\curv] =&\int_{\mathcal M_{\U}} \!\! \difffour{x} \ a^3 \e \left[ \dot{\curv}^2 - \frac{1}{a^2} \left(\pa \curv\right)^2 \right] \\
		&+ \int_{\partial \mathcal M_{\U}} \!\! \diffcubed{x} \ a^3 \e \left(-\frac{\eta}{2} \curv^2 \dot{\curv} - \frac{1}{H} \curv \dot{\curv}^2\right)
		\numberthis
		\end{align*}
		where the first line is the quadratic action and the second line is the cubic action, $\mathcal M_{\U}$ denotes the bulk of the spacetime $\vec{x} \in (-\vec{\infty}, +\vec{\infty} )$ and $t \in (0, t^{\U}_{\text{end}})$, and $\partial \mathcal M_{\U}$ denotes the temporal boundary of $\mathcal M_{\U}$ at $t^{\U}_{\text{end}}$. The cubic action here can be obtained from Eq.~\eqref{eq:cubicLagrangianSqueezed} by 
		eliminating $\dot\eta$- and $\e$-suppressed terms, which are both driven to zero in USR.

		Note that plugging the cubic portion of this action into the in-in formula Eq.~\eqref{eq:inin} with the modefunctions above yields the same result as Eq.~\eqref{eq:analyticresult}.

		Let us transform the action~\eqref{eq:S23c} in two ways.
		Transforming the action~\eqref{eq:S23c} using Eq.~\eqref{eq:timetransform} as a field redefinition alone gives
		\begin{align*}
		\label{eq:zetaNU}
		S[\curv_N] =&\int_{\mathcal M_{\U}}\!\! \difffour{x} \ \L_2 [\curv_N] 
	 	+ \int_{\partial \mathcal M_{\U}}\!\! \diffcubed{x} \ (2 a^3 \e \dot{\curv}_N f) 
	 	\\&+ \int_{\partial \mathcal M_{\U}}\!\! \diffcubed{x} \ a^3 \e \left(-\frac{\eta}{2} \curv_N^2 \dot{\curv}_N - \frac{1}{H} \curv_N \dot{\curv}_N^2\right), \numberthis
		\end{align*}
		where we have integrated by parts and used the equation of motion from the quadratic action 
		$\L_2[\curv_N]$ defined in \eqref{eq:L2}. 

		After substituting Eq.~\eqref{eq:nonlinearterm} for the $f$ term associated with the nonlinear field
		redefinition, we obtain 
		\begin{equation}
		\label{eq:zetaNU2}
		S[\curv_N] =\int_{\mathcal M_{\U}}\!\! \difffour{x} \ \L_2 [\curv_N] 
		+ \int_{\partial \mathcal M_{\U}}\!\! \diffcubed{x} \ \frac{a^3 \e}{H} \curv_N \dot{\curv}_N^2 ,
		\end{equation}
		where we see that a cubic boundary interaction remains in the action. 
		Unlike in the ordinary SR case, it contributes significantly in the USR case since $\zeta_N\propto a^3$ outside the horizon.

		Therefore, if we had used Eq.~\eqref{eq:zetaNU2} for the cubic interactions in the in-in formula, we would have computed
		\begin{equation}
		\langle \curv_N \curv_N \curv_N \rangle \neq 0,
		\end{equation}
		which is inconsistent with what we expected.

		In contrast, if we interpret Eq.~\eqref{eq:timetransform} as a gauge transformation, then
		we also have to apply the transformation of the temporal boundary 
		$\partial \mathcal M_{\U} = \partial \mathcal M_{\SF} + {\cal O}(\curv_N)$ to the action~\eqref{eq:zetaNU2}. 
		For the intrinsically cubic terms, 
		this transformation yields a higher-order correction, 
		but for the quadratic term it produces a cubic boundary interaction 
		\begin{eqnarray}
		\int_{\mathcal M_{\U}}\!\! \difffour{x} \ \L_2 [\curv_N] &=& \int_{\mathcal M_{\SF}}\!\! \difffour{x} \ \L_2 [\curv_N] \\ 
		&&\quad - \int_{\partial \mathcal M_{\SF}}\!\! \diffcubed{x} \ \frac{\curv_N}{H} \L_2 [\curv_N]\, , \nonumber
		\end{eqnarray}
		which precisely cancels the remaining cubic order term after ignoring higher-order terms of $\e$ and $k/(aH)$.
		We thus end up with the action of free field form
		\begin{equation}
		S[\curv_N] =\int_{\mathcal M_{\SF}}\!\! \difffour{x} \ a^3 \e \left[ \dot{\curv}_N^2 - \frac{1}{a^2} \left(\pa \curv_N\right)^2 \right],
		\end{equation}
		up to the cubic order of $\zeta$.
		Since there is no cubic interaction for $\curv_N$, we have 
		\begin{equation}
		\langle \curv_N \curv_N \curv_N \rangle = 0.
		\end{equation}
		This is consistent with the intuition that a flat potential for the scalar field produces no interaction terms.

		Therefore, the transformation \eqref{eq:timetransform} is better viewed as a gauge transformation than a field redefinition.  
		In practice, realistic inflationary models have an end of the USR stage. We therefore can always choose to evaluate the bispectrum after all relevant modes have frozen out and in this case the $\curv \dot{\curv}^2$ boundary term and the subtleties about evaluation times become irrelevant. This is the approach which is described in App.~\ref{app:numerics}.

\bibliographystyle{apsrev4-1}
\bibliography{../references.bib}

\begin{thebibliography}{66}%
\makeatletter
\providecommand \@ifxundefined [1]{%
 \@ifx{#1\undefined}
}%
\providecommand \@ifnum [1]{%
 \ifnum #1\expandafter \@firstoftwo
 \else \expandafter \@secondoftwo
 \fi
}%
\providecommand \@ifx [1]{%
 \ifx #1\expandafter \@firstoftwo
 \else \expandafter \@secondoftwo
 \fi
}%
\providecommand \natexlab [1]{#1}%
\providecommand \enquote  [1]{``#1''}%
\providecommand \bibnamefont  [1]{#1}%
\providecommand \bibfnamefont [1]{#1}%
\providecommand \citenamefont [1]{#1}%
\providecommand \href@noop [0]{\@secondoftwo}%
\providecommand \href [0]{\begingroup \@sanitize@url \@href}%
\providecommand \@href[1]{\@@startlink{#1}\@@href}%
\providecommand \@@href[1]{\endgroup#1\@@endlink}%
\providecommand \@sanitize@url [0]{\catcode `\\12\catcode `\$12\catcode
  `\&12\catcode `\#12\catcode `\^12\catcode `\_12\catcode `\%12\relax}%
\providecommand \@@startlink[1]{}%
\providecommand \@@endlink[0]{}%
\providecommand \url  [0]{\begingroup\@sanitize@url \@url }%
\providecommand \@url [1]{\endgroup\@href {#1}{\urlprefix }}%
\providecommand \urlprefix  [0]{URL }%
\providecommand \Eprint [0]{\href }%
\providecommand \doibase [0]{http://dx.doi.org/}%
\providecommand \selectlanguage [0]{\@gobble}%
\providecommand \bibinfo  [0]{\@secondoftwo}%
\providecommand \bibfield  [0]{\@secondoftwo}%
\providecommand \translation [1]{[#1]}%
\providecommand \BibitemOpen [0]{}%
\providecommand \bibitemStop [0]{}%
\providecommand \bibitemNoStop [0]{.\EOS\space}%
\providecommand \EOS [0]{\spacefactor3000\relax}%
\providecommand \BibitemShut  [1]{\csname bibitem#1\endcsname}%
\let\auto@bib@innerbib\@empty
\bibitem [{\citenamefont {{Zel'dovich}}\ and\ \citenamefont
  {{Novikov}}(1967)}]{Zeldovich:1967aaa}%
  \BibitemOpen
  \bibfield  {author} {\bibinfo {author} {\bibfnamefont {Y.~B.}\ \bibnamefont
  {{Zel'dovich}}}\ and\ \bibinfo {author} {\bibfnamefont {I.~D.}\ \bibnamefont
  {{Novikov}}},\ }\href {http://adsabs.harvard.edu/abs/1967SvA....10..602Z}
  {\bibfield  {journal} {\bibinfo  {journal} {Sov. Astron.}\ }\textbf {\bibinfo
  {volume} {10}},\ \bibinfo {pages} {602} (\bibinfo {year} {1967})}\BibitemShut
  {NoStop}%
\bibitem [{\citenamefont {{Carr}}\ and\ \citenamefont
  {{Hawking}}(1974)}]{Carr:1974}%
  \BibitemOpen
  \bibfield  {author} {\bibinfo {author} {\bibfnamefont {B.~J.}\ \bibnamefont
  {{Carr}}}\ and\ \bibinfo {author} {\bibfnamefont {S.~W.}\ \bibnamefont
  {{Hawking}}},\ }\href {\doibase 10.1093/mnras/168.2.399} {\bibfield
  {journal} {\bibinfo  {journal} {Mon. Not. Roy. Astron. Soc.}\ }\textbf
  {\bibinfo {volume} {168}},\ \bibinfo {pages} {399} (\bibinfo {year}
  {1974})}\BibitemShut {NoStop}%
\bibitem [{\citenamefont {{Carr}}(1975)}]{Carr:1975aaa}%
  \BibitemOpen
  \bibfield  {author} {\bibinfo {author} {\bibfnamefont {B.~J.}\ \bibnamefont
  {{Carr}}},\ }\href {\doibase 10.1086/153853} {\bibfield  {journal} {\bibinfo
  {journal} {\apj}\ }\textbf {\bibinfo {volume} {201}},\ \bibinfo {pages} {1}
  (\bibinfo {year} {1975})}\BibitemShut {NoStop}%
\bibitem [{\citenamefont {{Meszaros}}(1974)}]{Meszaros:1974aaa}%
  \BibitemOpen
  \bibfield  {author} {\bibinfo {author} {\bibfnamefont {P.}~\bibnamefont
  {{Meszaros}}},\ }\href {http://adsabs.harvard.edu/abs/1974A%26A....37..225M}
  {\bibfield  {journal} {\bibinfo  {journal} {Astron. Astrophys.}\ }\textbf
  {\bibinfo {volume} {37}},\ \bibinfo {pages} {225} (\bibinfo {year}
  {1974})}\BibitemShut {NoStop}%
\bibitem [{\citenamefont {{Chapline}}(1975)}]{Chapline:1975aaa}%
  \BibitemOpen
  \bibfield  {author} {\bibinfo {author} {\bibfnamefont {G.~F.}\ \bibnamefont
  {{Chapline}}},\ }\href {\doibase 10.1038/253251a0} {\bibfield  {journal}
  {\bibinfo  {journal} {\nat}\ }\textbf {\bibinfo {volume} {253}},\ \bibinfo
  {pages} {251} (\bibinfo {year} {1975})}\BibitemShut {NoStop}%
\bibitem [{\citenamefont {Carr}\ \emph {et~al.}(2016)\citenamefont {Carr},
  \citenamefont {K{\"u}hnel},\ and\ \citenamefont {Sandstad}}]{Carr:2016drx}%
  \BibitemOpen
  \bibfield  {author} {\bibinfo {author} {\bibfnamefont {B.}~\bibnamefont
  {Carr}}, \bibinfo {author} {\bibfnamefont {F.}~\bibnamefont {K{\"u}hnel}}, \
  and\ \bibinfo {author} {\bibfnamefont {M.}~\bibnamefont {Sandstad}},\ }\href
  {\doibase 10.1103/PhysRevD.94.083504} {\bibfield  {journal} {\bibinfo
  {journal} {Phys. Rev.}\ }\textbf {\bibinfo {volume} {D94}},\ \bibinfo {pages}
  {083504} (\bibinfo {year} {2016})},\ \Eprint
  {http://arxiv.org/abs/1607.06077} {arXiv:1607.06077 [astro-ph.CO]}
  \BibitemShut {NoStop}%
\bibitem [{\citenamefont {Niikura}\ \emph {et~al.}()\citenamefont {Niikura}
  \emph {et~al.}}]{Niikura:2017zjd}%
  \BibitemOpen
  \bibfield  {author} {\bibinfo {author} {\bibfnamefont {H.}~\bibnamefont
  {Niikura}} \emph {et~al.},\ }\href@noop {} {\ }\Eprint
  {http://arxiv.org/abs/1701.02151} {arXiv:1701.02151 [astro-ph.CO]}
  \BibitemShut {NoStop}%
\bibitem [{\citenamefont {Carr}\ \emph {et~al.}(2017)\citenamefont {Carr},
  \citenamefont {Raidal}, \citenamefont {Tenkanen}, \citenamefont {Vaskonen},\
  and\ \citenamefont {Veerm{\"a}e}}]{Carr:2017jsz}%
  \BibitemOpen
  \bibfield  {author} {\bibinfo {author} {\bibfnamefont {B.}~\bibnamefont
  {Carr}}, \bibinfo {author} {\bibfnamefont {M.}~\bibnamefont {Raidal}},
  \bibinfo {author} {\bibfnamefont {T.}~\bibnamefont {Tenkanen}}, \bibinfo
  {author} {\bibfnamefont {V.}~\bibnamefont {Vaskonen}}, \ and\ \bibinfo
  {author} {\bibfnamefont {H.}~\bibnamefont {Veerm{\"a}e}},\ }\href {\doibase
  10.1103/PhysRevD.96.023514} {\bibfield  {journal} {\bibinfo  {journal} {Phys.
  Rev.}\ }\textbf {\bibinfo {volume} {D96}},\ \bibinfo {pages} {023514}
  (\bibinfo {year} {2017})},\ \Eprint {http://arxiv.org/abs/1705.05567}
  {arXiv:1705.05567 [astro-ph.CO]} \BibitemShut {NoStop}%
\bibitem [{\citenamefont {K{\"u}hnel}\ and\ \citenamefont
  {Freese}(2017)}]{Kuhnel:2017pwq}%
  \BibitemOpen
  \bibfield  {author} {\bibinfo {author} {\bibfnamefont {F.}~\bibnamefont
  {K{\"u}hnel}}\ and\ \bibinfo {author} {\bibfnamefont {K.}~\bibnamefont
  {Freese}},\ }\href {\doibase 10.1103/PhysRevD.95.083508} {\bibfield
  {journal} {\bibinfo  {journal} {Phys. Rev.}\ }\textbf {\bibinfo {volume}
  {D95}},\ \bibinfo {pages} {083508} (\bibinfo {year} {2017})},\ \Eprint
  {http://arxiv.org/abs/1701.07223} {arXiv:1701.07223 [astro-ph.CO]}
  \BibitemShut {NoStop}%
\bibitem [{\citenamefont {Capela}\ \emph {et~al.}(2013)\citenamefont {Capela},
  \citenamefont {Pshirkov},\ and\ \citenamefont {Tinyakov}}]{Capela:2013yf}%
  \BibitemOpen
  \bibfield  {author} {\bibinfo {author} {\bibfnamefont {F.}~\bibnamefont
  {Capela}}, \bibinfo {author} {\bibfnamefont {M.}~\bibnamefont {Pshirkov}}, \
  and\ \bibinfo {author} {\bibfnamefont {P.}~\bibnamefont {Tinyakov}},\ }\href
  {\doibase 10.1103/PhysRevD.87.123524} {\bibfield  {journal} {\bibinfo
  {journal} {Phys. Rev.}\ }\textbf {\bibinfo {volume} {D87}},\ \bibinfo {pages}
  {123524} (\bibinfo {year} {2013})},\ \Eprint {http://arxiv.org/abs/1301.4984}
  {arXiv:1301.4984 [astro-ph.CO]} \BibitemShut {NoStop}%
\bibitem [{\citenamefont {{Lane}}\ \emph {et~al.}(2009)\citenamefont {{Lane}},
  \citenamefont {{Kiss}}, \citenamefont {{Lewis}}, \citenamefont {{Ibata}},
  \citenamefont {{Siebert}}, \citenamefont {{Bedding}},\ and\ \citenamefont
  {{Sz{\'e}kely}}}]{Lane:2009aa}%
  \BibitemOpen
  \bibfield  {author} {\bibinfo {author} {\bibfnamefont {R.~R.}\ \bibnamefont
  {{Lane}}}, \bibinfo {author} {\bibfnamefont {L.~L.}\ \bibnamefont {{Kiss}}},
  \bibinfo {author} {\bibfnamefont {G.~F.}\ \bibnamefont {{Lewis}}}, \bibinfo
  {author} {\bibfnamefont {R.~A.}\ \bibnamefont {{Ibata}}}, \bibinfo {author}
  {\bibfnamefont {A.}~\bibnamefont {{Siebert}}}, \bibinfo {author}
  {\bibfnamefont {T.~R.}\ \bibnamefont {{Bedding}}}, \ and\ \bibinfo {author}
  {\bibfnamefont {P.}~\bibnamefont {{Sz{\'e}kely}}},\ }\href {\doibase
  10.1111/j.1365-2966.2009.15505.x} {\bibfield  {journal} {\bibinfo  {journal}
  {Mon. Not. Roy. Astron. Soc.}\ }\textbf {\bibinfo {volume} {400}},\ \bibinfo
  {pages} {917} (\bibinfo {year} {2009})},\ \Eprint
  {http://arxiv.org/abs/0908.0770} {arXiv:0908.0770 [astro-ph.GA]} \BibitemShut
  {NoStop}%
\bibitem [{\citenamefont {Bird}\ \emph {et~al.}(2016)\citenamefont {Bird},
  \citenamefont {Cholis}, \citenamefont {Mu{\~n}oz}, \citenamefont
  {Ali-Ha{\"i}moud}, \citenamefont {Kamionkowski}, \citenamefont {Kovetz},
  \citenamefont {Raccanelli},\ and\ \citenamefont {Riess}}]{Bird:2016dcv}%
  \BibitemOpen
  \bibfield  {author} {\bibinfo {author} {\bibfnamefont {S.}~\bibnamefont
  {Bird}}, \bibinfo {author} {\bibfnamefont {I.}~\bibnamefont {Cholis}},
  \bibinfo {author} {\bibfnamefont {J.~B.}\ \bibnamefont {Mu{\~n}oz}}, \bibinfo
  {author} {\bibfnamefont {Y.}~\bibnamefont {Ali-Ha{\"i}moud}}, \bibinfo
  {author} {\bibfnamefont {M.}~\bibnamefont {Kamionkowski}}, \bibinfo {author}
  {\bibfnamefont {E.~D.}\ \bibnamefont {Kovetz}}, \bibinfo {author}
  {\bibfnamefont {A.}~\bibnamefont {Raccanelli}}, \ and\ \bibinfo {author}
  {\bibfnamefont {A.~G.}\ \bibnamefont {Riess}},\ }\href {\doibase
  10.1103/PhysRevLett.116.201301} {\bibfield  {journal} {\bibinfo  {journal}
  {Phys. Rev. Lett.}\ }\textbf {\bibinfo {volume} {116}},\ \bibinfo {pages}
  {201301} (\bibinfo {year} {2016})},\ \Eprint
  {http://arxiv.org/abs/1603.00464} {arXiv:1603.00464 [astro-ph.CO]}
  \BibitemShut {NoStop}%
\bibitem [{\citenamefont {Sasaki}\ \emph {et~al.}(2016)\citenamefont {Sasaki},
  \citenamefont {Suyama}, \citenamefont {Tanaka},\ and\ \citenamefont
  {Yokoyama}}]{Sasaki:2016jop}%
  \BibitemOpen
  \bibfield  {author} {\bibinfo {author} {\bibfnamefont {M.}~\bibnamefont
  {Sasaki}}, \bibinfo {author} {\bibfnamefont {T.}~\bibnamefont {Suyama}},
  \bibinfo {author} {\bibfnamefont {T.}~\bibnamefont {Tanaka}}, \ and\ \bibinfo
  {author} {\bibfnamefont {S.}~\bibnamefont {Yokoyama}},\ }\href {\doibase
  10.1103/PhysRevLett.117.061101} {\bibfield  {journal} {\bibinfo  {journal}
  {Phys. Rev. Lett.}\ }\textbf {\bibinfo {volume} {117}},\ \bibinfo {pages}
  {061101} (\bibinfo {year} {2016})},\ \bibinfo {note} {[Erratum:
  \href{http://dx.doi.org/10.1103/PhysRevLett.121.059901}{Phys. Rev. Lett.
  \textbf{121}, 059901 (2018)}]},\ \Eprint {http://arxiv.org/abs/1603.08338}
  {arXiv:1603.08338 [astro-ph.CO]} \BibitemShut {NoStop}%
\bibitem [{\citenamefont {Zumalac\'arregui}\ and\ \citenamefont
  {Seljak}(2018)}]{Zumalacarregui:2017qqd}%
  \BibitemOpen
  \bibfield  {author} {\bibinfo {author} {\bibfnamefont {M.}~\bibnamefont
  {Zumalac\'arregui}}\ and\ \bibinfo {author} {\bibfnamefont {U.}~\bibnamefont
  {Seljak}},\ }\href {\doibase 10.1103/PhysRevLett.121.141101} {\bibfield
  {journal} {\bibinfo  {journal} {Phys. Rev. Lett.}\ }\textbf {\bibinfo
  {volume} {121}},\ \bibinfo {pages} {141101} (\bibinfo {year} {2018})},\
  \Eprint {http://arxiv.org/abs/1712.02240} {arXiv:1712.02240 [astro-ph.CO]}
  \BibitemShut {NoStop}%
\bibitem [{\citenamefont {Sasaki}\ \emph {et~al.}(2018)\citenamefont {Sasaki},
  \citenamefont {Suyama}, \citenamefont {Tanaka},\ and\ \citenamefont
  {Yokoyama}}]{Sasaki:2018dmp}%
  \BibitemOpen
  \bibfield  {author} {\bibinfo {author} {\bibfnamefont {M.}~\bibnamefont
  {Sasaki}}, \bibinfo {author} {\bibfnamefont {T.}~\bibnamefont {Suyama}},
  \bibinfo {author} {\bibfnamefont {T.}~\bibnamefont {Tanaka}}, \ and\ \bibinfo
  {author} {\bibfnamefont {S.}~\bibnamefont {Yokoyama}},\ }\href {\doibase
  10.1088/1361-6382/aaa7b4} {\bibfield  {journal} {\bibinfo  {journal} {Class.
  Quant. Grav.}\ }\textbf {\bibinfo {volume} {35}},\ \bibinfo {pages} {063001}
  (\bibinfo {year} {2018})},\ \Eprint {http://arxiv.org/abs/1801.05235}
  {arXiv:1801.05235 [astro-ph.CO]} \BibitemShut {NoStop}%
\bibitem [{\citenamefont {{LIGO Scientific and Virgo Collaborations
  (2018)}}()}]{LIGOScientific:2018mvr}%
  \BibitemOpen
  \bibfield  {author} {\bibinfo {author} {\bibnamefont {{LIGO Scientific and
  Virgo Collaborations (2018)}}},\ }\href@noop {} {\ }\Eprint
  {http://arxiv.org/abs/1811.12907} {arXiv:1811.12907 [astro-ph.HE]}
  \BibitemShut {NoStop}%
\bibitem [{\citenamefont {Motohashi}\ and\ \citenamefont
  {Hu}(2017)}]{Motohashi:2017kbs}%
  \BibitemOpen
  \bibfield  {author} {\bibinfo {author} {\bibfnamefont {H.}~\bibnamefont
  {Motohashi}}\ and\ \bibinfo {author} {\bibfnamefont {W.}~\bibnamefont {Hu}},\
  }\href {\doibase 10.1103/PhysRevD.96.063503} {\bibfield  {journal} {\bibinfo
  {journal} {Phys. Rev.}\ }\textbf {\bibinfo {volume} {D96}},\ \bibinfo {pages}
  {063503} (\bibinfo {year} {2017})},\ \Eprint
  {http://arxiv.org/abs/1706.06784} {arXiv:1706.06784 [astro-ph.CO]}
  \BibitemShut {NoStop}%
\bibitem [{\citenamefont {Ballesteros}\ and\ \citenamefont
  {Taoso}(2018)}]{Ballesteros:2017fsr}%
  \BibitemOpen
  \bibfield  {author} {\bibinfo {author} {\bibfnamefont {G.}~\bibnamefont
  {Ballesteros}}\ and\ \bibinfo {author} {\bibfnamefont {M.}~\bibnamefont
  {Taoso}},\ }\href {\doibase 10.1103/PhysRevD.97.023501} {\bibfield  {journal}
  {\bibinfo  {journal} {Phys. Rev.}\ }\textbf {\bibinfo {volume} {D97}},\
  \bibinfo {pages} {023501} (\bibinfo {year} {2018})},\ \Eprint
  {http://arxiv.org/abs/1709.05565} {arXiv:1709.05565 [hep-ph]} \BibitemShut
  {NoStop}%
\bibitem [{\citenamefont {Kawasaki}\ \emph {et~al.}(2016)\citenamefont
  {Kawasaki}, \citenamefont {Kusenko}, \citenamefont {Tada},\ and\
  \citenamefont {Yanagida}}]{Kawasaki:2016pql}%
  \BibitemOpen
  \bibfield  {author} {\bibinfo {author} {\bibfnamefont {M.}~\bibnamefont
  {Kawasaki}}, \bibinfo {author} {\bibfnamefont {A.}~\bibnamefont {Kusenko}},
  \bibinfo {author} {\bibfnamefont {Y.}~\bibnamefont {Tada}}, \ and\ \bibinfo
  {author} {\bibfnamefont {T.~T.}\ \bibnamefont {Yanagida}},\ }\href {\doibase
  10.1103/PhysRevD.94.083523} {\bibfield  {journal} {\bibinfo  {journal} {Phys.
  Rev.}\ }\textbf {\bibinfo {volume} {D94}},\ \bibinfo {pages} {083523}
  (\bibinfo {year} {2016})},\ \Eprint {http://arxiv.org/abs/1606.07631}
  {arXiv:1606.07631 [astro-ph.CO]} \BibitemShut {NoStop}%
\bibitem [{\citenamefont {Garc\'ia-Bellido}\ and\ \citenamefont
  {Ruiz~Morales}(2017)}]{Garcia-Bellido:2017mdw}%
  \BibitemOpen
  \bibfield  {author} {\bibinfo {author} {\bibfnamefont {J.}~\bibnamefont
  {Garc\'ia-Bellido}}\ and\ \bibinfo {author} {\bibfnamefont {E.}~\bibnamefont
  {Ruiz~Morales}},\ }\href {\doibase 10.1016/j.dark.2017.09.007} {\bibfield
  {journal} {\bibinfo  {journal} {Phys. Dark Univ.}\ }\textbf {\bibinfo
  {volume} {18}},\ \bibinfo {pages} {47} (\bibinfo {year} {2017})},\ \Eprint
  {http://arxiv.org/abs/1702.03901} {arXiv:1702.03901 [astro-ph.CO]}
  \BibitemShut {NoStop}%
\bibitem [{\citenamefont {Kamenshchik}\ \emph {et~al.}()\citenamefont
  {Kamenshchik}, \citenamefont {Tronconi}, \citenamefont {Vardanyan},\ and\
  \citenamefont {Venturi}}]{Kamenshchik:2018sig}%
  \BibitemOpen
  \bibfield  {author} {\bibinfo {author} {\bibfnamefont {A.~Y.}\ \bibnamefont
  {Kamenshchik}}, \bibinfo {author} {\bibfnamefont {A.}~\bibnamefont
  {Tronconi}}, \bibinfo {author} {\bibfnamefont {T.}~\bibnamefont {Vardanyan}},
  \ and\ \bibinfo {author} {\bibfnamefont {G.}~\bibnamefont {Venturi}},\
  }\href@noop {} {\ }\Eprint {http://arxiv.org/abs/1812.02547}
  {arXiv:1812.02547 [gr-qc]} \BibitemShut {NoStop}%
\bibitem [{\citenamefont {Suyama}\ and\ \citenamefont
  {Yokoyama}(2011)}]{Suyama:2011pu}%
  \BibitemOpen
  \bibfield  {author} {\bibinfo {author} {\bibfnamefont {T.}~\bibnamefont
  {Suyama}}\ and\ \bibinfo {author} {\bibfnamefont {J.}~\bibnamefont
  {Yokoyama}},\ }\href {\doibase 10.1103/PhysRevD.84.083511} {\bibfield
  {journal} {\bibinfo  {journal} {Phys. Rev.}\ }\textbf {\bibinfo {volume}
  {D84}},\ \bibinfo {pages} {083511} (\bibinfo {year} {2011})},\ \Eprint
  {http://arxiv.org/abs/1106.5983} {arXiv:1106.5983 [astro-ph.CO]} \BibitemShut
  {NoStop}%
\bibitem [{\citenamefont {Kawasaki}\ \emph {et~al.}(2013)\citenamefont
  {Kawasaki}, \citenamefont {Kitajima},\ and\ \citenamefont
  {Yanagida}}]{Kawasaki:2012wr}%
  \BibitemOpen
  \bibfield  {author} {\bibinfo {author} {\bibfnamefont {M.}~\bibnamefont
  {Kawasaki}}, \bibinfo {author} {\bibfnamefont {N.}~\bibnamefont {Kitajima}},
  \ and\ \bibinfo {author} {\bibfnamefont {T.~T.}\ \bibnamefont {Yanagida}},\
  }\href {\doibase 10.1103/PhysRevD.87.063519} {\bibfield  {journal} {\bibinfo
  {journal} {Phys. Rev.}\ }\textbf {\bibinfo {volume} {D87}},\ \bibinfo {pages}
  {063519} (\bibinfo {year} {2013})},\ \Eprint {http://arxiv.org/abs/1207.2550}
  {arXiv:1207.2550 [hep-ph]} \BibitemShut {NoStop}%
\bibitem [{\citenamefont {Chen}\ \emph {et~al.}(2017)\citenamefont {Chen},
  \citenamefont {Liu}, \citenamefont {Xu},\ and\ \citenamefont
  {Cai}}]{Chen:2016kjx}%
  \BibitemOpen
  \bibfield  {author} {\bibinfo {author} {\bibfnamefont {J.-W.}\ \bibnamefont
  {Chen}}, \bibinfo {author} {\bibfnamefont {J.}~\bibnamefont {Liu}}, \bibinfo
  {author} {\bibfnamefont {H.-L.}\ \bibnamefont {Xu}}, \ and\ \bibinfo {author}
  {\bibfnamefont {Y.-F.}\ \bibnamefont {Cai}},\ }\href {\doibase
  10.1016/j.physletb.2017.03.036} {\bibfield  {journal} {\bibinfo  {journal}
  {Phys. Lett.}\ }\textbf {\bibinfo {volume} {B769}},\ \bibinfo {pages} {561}
  (\bibinfo {year} {2017})},\ \Eprint {http://arxiv.org/abs/1609.02571}
  {arXiv:1609.02571 [gr-qc]} \BibitemShut {NoStop}%
\bibitem [{\citenamefont {Young}\ and\ \citenamefont
  {Byrnes}(2013)}]{Young:2013oia}%
  \BibitemOpen
  \bibfield  {author} {\bibinfo {author} {\bibfnamefont {S.}~\bibnamefont
  {Young}}\ and\ \bibinfo {author} {\bibfnamefont {C.~T.}\ \bibnamefont
  {Byrnes}},\ }\href {\doibase 10.1088/1475-7516/2013/08/052} {\bibfield
  {journal} {\bibinfo  {journal} {JCAP}\ }\textbf {\bibinfo {volume} {1308}},\
  \bibinfo {pages} {052} (\bibinfo {year} {2013})},\ \Eprint
  {http://arxiv.org/abs/1307.4995} {arXiv:1307.4995 [astro-ph.CO]} \BibitemShut
  {NoStop}%
\bibitem [{\citenamefont {Young}\ \emph {et~al.}(2014)\citenamefont {Young},
  \citenamefont {Byrnes},\ and\ \citenamefont {Sasaki}}]{Young:2014ana}%
  \BibitemOpen
  \bibfield  {author} {\bibinfo {author} {\bibfnamefont {S.}~\bibnamefont
  {Young}}, \bibinfo {author} {\bibfnamefont {C.~T.}\ \bibnamefont {Byrnes}}, \
  and\ \bibinfo {author} {\bibfnamefont {M.}~\bibnamefont {Sasaki}},\ }\href
  {\doibase 10.1088/1475-7516/2014/07/045} {\bibfield  {journal} {\bibinfo
  {journal} {JCAP}\ }\textbf {\bibinfo {volume} {1407}},\ \bibinfo {pages}
  {045} (\bibinfo {year} {2014})},\ \Eprint {http://arxiv.org/abs/1405.7023}
  {arXiv:1405.7023 [gr-qc]} \BibitemShut {NoStop}%
\bibitem [{\citenamefont {Franciolini}\ \emph {et~al.}(2018)\citenamefont
  {Franciolini}, \citenamefont {Kehagias}, \citenamefont {Matarrese},\ and\
  \citenamefont {Riotto}}]{Franciolini:2018vbk}%
  \BibitemOpen
  \bibfield  {author} {\bibinfo {author} {\bibfnamefont {G.}~\bibnamefont
  {Franciolini}}, \bibinfo {author} {\bibfnamefont {A.}~\bibnamefont
  {Kehagias}}, \bibinfo {author} {\bibfnamefont {S.}~\bibnamefont {Matarrese}},
  \ and\ \bibinfo {author} {\bibfnamefont {A.}~\bibnamefont {Riotto}},\ }\href
  {\doibase 10.1088/1475-7516/2018/03/016} {\bibfield  {journal} {\bibinfo
  {journal} {JCAP}\ }\textbf {\bibinfo {volume} {1803}},\ \bibinfo {pages}
  {016} (\bibinfo {year} {2018})},\ \Eprint {http://arxiv.org/abs/1801.09415}
  {arXiv:1801.09415 [astro-ph.CO]} \BibitemShut {NoStop}%
\bibitem [{\citenamefont {Germani}\ and\ \citenamefont
  {Musco}()}]{Germani:2018jgr}%
  \BibitemOpen
  \bibfield  {author} {\bibinfo {author} {\bibfnamefont {C.}~\bibnamefont
  {Germani}}\ and\ \bibinfo {author} {\bibfnamefont {I.}~\bibnamefont
  {Musco}},\ }\href@noop {} {\ }\Eprint {http://arxiv.org/abs/1805.04087}
  {arXiv:1805.04087 [astro-ph.CO]} \BibitemShut {NoStop}%
\bibitem [{\citenamefont {Kinney}(2005)}]{Kinney:2005vj}%
  \BibitemOpen
  \bibfield  {author} {\bibinfo {author} {\bibfnamefont {W.~H.}\ \bibnamefont
  {Kinney}},\ }\href {\doibase 10.1103/PhysRevD.72.023515} {\bibfield
  {journal} {\bibinfo  {journal} {Phys. Rev.}\ }\textbf {\bibinfo {volume}
  {D72}},\ \bibinfo {pages} {023515} (\bibinfo {year} {2005})},\ \Eprint
  {http://arxiv.org/abs/gr-qc/0503017} {arXiv:gr-qc/0503017 [gr-qc]}
  \BibitemShut {NoStop}%
\bibitem [{\citenamefont {Starobinsky}(1985)}]{Starobinsky:1985aa}%
  \BibitemOpen
  \bibfield  {author} {\bibinfo {author} {\bibfnamefont {A.~A.}\ \bibnamefont
  {Starobinsky}},\ }\href
  {http://www.jetpletters.ac.ru/ps/1419/article_21563.shtml} {\bibfield
  {journal} {\bibinfo  {journal} {JETP Lett.}\ }\textbf {\bibinfo {volume}
  {42}},\ \bibinfo {pages} {152} (\bibinfo {year} {1985})},\ \bibinfo {note}
  {\href{http://www.jetpletters.ac.ru/ps/98/article_1729.shtml}{Pis'ma v ZhETF.
  42, 124 (1985)}}\BibitemShut {NoStop}%
\bibitem [{\citenamefont {Salopek}\ and\ \citenamefont
  {Bond}(1990)}]{Salopek:1990jq}%
  \BibitemOpen
  \bibfield  {author} {\bibinfo {author} {\bibfnamefont {D.~S.}\ \bibnamefont
  {Salopek}}\ and\ \bibinfo {author} {\bibfnamefont {J.~R.}\ \bibnamefont
  {Bond}},\ }\href {\doibase 10.1103/PhysRevD.42.3936} {\bibfield  {journal}
  {\bibinfo  {journal} {Phys. Rev.}\ }\textbf {\bibinfo {volume} {D42}},\
  \bibinfo {pages} {3936} (\bibinfo {year} {1990})}\BibitemShut {NoStop}%
\bibitem [{\citenamefont {Sasaki}\ and\ \citenamefont
  {Stewart}(1996)}]{Sasaki:1995aw}%
  \BibitemOpen
  \bibfield  {author} {\bibinfo {author} {\bibfnamefont {M.}~\bibnamefont
  {Sasaki}}\ and\ \bibinfo {author} {\bibfnamefont {E.~D.}\ \bibnamefont
  {Stewart}},\ }\href {\doibase 10.1143/PTP.95.71} {\bibfield  {journal}
  {\bibinfo  {journal} {Prog. Theor. Phys.}\ }\textbf {\bibinfo {volume}
  {95}},\ \bibinfo {pages} {71} (\bibinfo {year} {1996})},\ \Eprint
  {http://arxiv.org/abs/astro-ph/9507001} {arXiv:astro-ph/9507001 [astro-ph]}
  \BibitemShut {NoStop}%
\bibitem [{\citenamefont {Sugiyama}\ \emph {et~al.}(2013)\citenamefont
  {Sugiyama}, \citenamefont {Komatsu},\ and\ \citenamefont
  {Futamase}}]{Sugiyama:2012tj}%
  \BibitemOpen
  \bibfield  {author} {\bibinfo {author} {\bibfnamefont {N.~S.}\ \bibnamefont
  {Sugiyama}}, \bibinfo {author} {\bibfnamefont {E.}~\bibnamefont {Komatsu}}, \
  and\ \bibinfo {author} {\bibfnamefont {T.}~\bibnamefont {Futamase}},\ }\href
  {\doibase 10.1103/PhysRevD.87.023530} {\bibfield  {journal} {\bibinfo
  {journal} {Phys. Rev.}\ }\textbf {\bibinfo {volume} {D87}},\ \bibinfo {pages}
  {023530} (\bibinfo {year} {2013})},\ \Eprint {http://arxiv.org/abs/1208.1073}
  {arXiv:1208.1073 [gr-qc]} \BibitemShut {NoStop}%
\bibitem [{\citenamefont {Domenech}\ \emph {et~al.}(2017)\citenamefont
  {Domenech}, \citenamefont {Gong},\ and\ \citenamefont
  {Sasaki}}]{Domenech:2016zxn}%
  \BibitemOpen
  \bibfield  {author} {\bibinfo {author} {\bibfnamefont {G.}~\bibnamefont
  {Domenech}}, \bibinfo {author} {\bibfnamefont {J.-O.}\ \bibnamefont {Gong}},
  \ and\ \bibinfo {author} {\bibfnamefont {M.}~\bibnamefont {Sasaki}},\ }\href
  {\doibase 10.1016/j.physletb.2017.04.014} {\bibfield  {journal} {\bibinfo
  {journal} {Phys. Lett.}\ }\textbf {\bibinfo {volume} {B769}},\ \bibinfo
  {pages} {413} (\bibinfo {year} {2017})},\ \Eprint
  {http://arxiv.org/abs/1606.03343} {arXiv:1606.03343 [astro-ph.CO]}
  \BibitemShut {NoStop}%
\bibitem [{\citenamefont {Abolhasani}\ and\ \citenamefont
  {Sasaki}(2018)}]{Abolhasani:2018gyz}%
  \BibitemOpen
  \bibfield  {author} {\bibinfo {author} {\bibfnamefont {A.~A.}\ \bibnamefont
  {Abolhasani}}\ and\ \bibinfo {author} {\bibfnamefont {M.}~\bibnamefont
  {Sasaki}},\ }\href {\doibase 10.1088/1475-7516/2018/08/025} {\bibfield
  {journal} {\bibinfo  {journal} {JCAP}\ }\textbf {\bibinfo {volume} {1808}},\
  \bibinfo {pages} {025} (\bibinfo {year} {2018})},\ \Eprint
  {http://arxiv.org/abs/1805.11298} {arXiv:1805.11298 [astro-ph.CO]}
  \BibitemShut {NoStop}%
\bibitem [{\citenamefont {Maldacena}(2003)}]{Maldacena:2002vr}%
  \BibitemOpen
  \bibfield  {author} {\bibinfo {author} {\bibfnamefont {J.~M.}\ \bibnamefont
  {Maldacena}},\ }\href {\doibase 10.1088/1126-6708/2003/05/013} {\bibfield
  {journal} {\bibinfo  {journal} {JHEP}\ }\textbf {\bibinfo {volume} {05}},\
  \bibinfo {pages} {013} (\bibinfo {year} {2003})},\ \Eprint
  {http://arxiv.org/abs/astro-ph/0210603} {arXiv:astro-ph/0210603 [astro-ph]}
  \BibitemShut {NoStop}%
\bibitem [{\citenamefont {Cai}\ \emph {et~al.}(2018)\citenamefont {Cai},
  \citenamefont {Chen}, \citenamefont {Namjoo}, \citenamefont {Sasaki},
  \citenamefont {Wang},\ and\ \citenamefont {Wang}}]{Cai:2017bxr}%
  \BibitemOpen
  \bibfield  {author} {\bibinfo {author} {\bibfnamefont {Y.-F.}\ \bibnamefont
  {Cai}}, \bibinfo {author} {\bibfnamefont {X.}~\bibnamefont {Chen}}, \bibinfo
  {author} {\bibfnamefont {M.~H.}\ \bibnamefont {Namjoo}}, \bibinfo {author}
  {\bibfnamefont {M.}~\bibnamefont {Sasaki}}, \bibinfo {author} {\bibfnamefont
  {D.-G.}\ \bibnamefont {Wang}}, \ and\ \bibinfo {author} {\bibfnamefont
  {Z.}~\bibnamefont {Wang}},\ }\href {\doibase 10.1088/1475-7516/2018/05/012}
  {\bibfield  {journal} {\bibinfo  {journal} {JCAP}\ }\textbf {\bibinfo
  {volume} {1805}},\ \bibinfo {pages} {012} (\bibinfo {year} {2018})},\ \Eprint
  {http://arxiv.org/abs/1712.09998} {arXiv:1712.09998 [astro-ph.CO]}
  \BibitemShut {NoStop}%
\bibitem [{\citenamefont {Manasse}\ and\ \citenamefont
  {Misner}(1963)}]{Manasse:1963}%
  \BibitemOpen
  \bibfield  {author} {\bibinfo {author} {\bibfnamefont {F.~K.}\ \bibnamefont
  {Manasse}}\ and\ \bibinfo {author} {\bibfnamefont {C.~W.}\ \bibnamefont
  {Misner}},\ }\href {\doibase 10.1063/1.1724316} {\bibfield  {journal}
  {\bibinfo  {journal} {J. Math. Phys.}\ }\textbf {\bibinfo {volume} {4}},\
  \bibinfo {pages} {735} (\bibinfo {year} {1963})}\BibitemShut {NoStop}%
\bibitem [{\citenamefont {Senatore}\ and\ \citenamefont
  {Zaldarriaga}(2013)}]{Senatore:2012ya}%
  \BibitemOpen
  \bibfield  {author} {\bibinfo {author} {\bibfnamefont {L.}~\bibnamefont
  {Senatore}}\ and\ \bibinfo {author} {\bibfnamefont {M.}~\bibnamefont
  {Zaldarriaga}},\ }\href {\doibase 10.1007/JHEP09(2013)148} {\bibfield
  {journal} {\bibinfo  {journal} {JHEP}\ }\textbf {\bibinfo {volume} {09}},\
  \bibinfo {pages} {148} (\bibinfo {year} {2013})},\ \Eprint
  {http://arxiv.org/abs/1210.6048} {arXiv:1210.6048 [hep-th]} \BibitemShut
  {NoStop}%
\bibitem [{\citenamefont {Senatore}\ and\ \citenamefont
  {Zaldarriaga}(2012)}]{Senatore:2012wy}%
  \BibitemOpen
  \bibfield  {author} {\bibinfo {author} {\bibfnamefont {L.}~\bibnamefont
  {Senatore}}\ and\ \bibinfo {author} {\bibfnamefont {M.}~\bibnamefont
  {Zaldarriaga}},\ }\href {\doibase 10.1088/1475-7516/2012/08/001} {\bibfield
  {journal} {\bibinfo  {journal} {JCAP}\ }\textbf {\bibinfo {volume} {1208}},\
  \bibinfo {pages} {001} (\bibinfo {year} {2012})},\ \Eprint
  {http://arxiv.org/abs/1203.6884} {arXiv:1203.6884 [astro-ph.CO]} \BibitemShut
  {NoStop}%
\bibitem [{\citenamefont {Pajer}\ \emph {et~al.}(2013)\citenamefont {Pajer},
  \citenamefont {Schmidt},\ and\ \citenamefont {Zaldarriaga}}]{Pajer:2013ana}%
  \BibitemOpen
  \bibfield  {author} {\bibinfo {author} {\bibfnamefont {E.}~\bibnamefont
  {Pajer}}, \bibinfo {author} {\bibfnamefont {F.}~\bibnamefont {Schmidt}}, \
  and\ \bibinfo {author} {\bibfnamefont {M.}~\bibnamefont {Zaldarriaga}},\
  }\href {\doibase 10.1103/PhysRevD.88.083502} {\bibfield  {journal} {\bibinfo
  {journal} {Phys. Rev.}\ }\textbf {\bibinfo {volume} {D88}},\ \bibinfo {pages}
  {083502} (\bibinfo {year} {2013})},\ \Eprint {http://arxiv.org/abs/1305.0824}
  {arXiv:1305.0824 [astro-ph.CO]} \BibitemShut {NoStop}%
\bibitem [{\citenamefont {Namjoo}\ \emph {et~al.}(2013)\citenamefont {Namjoo},
  \citenamefont {Firouzjahi},\ and\ \citenamefont {Sasaki}}]{Namjoo:2012aa}%
  \BibitemOpen
  \bibfield  {author} {\bibinfo {author} {\bibfnamefont {M.~H.}\ \bibnamefont
  {Namjoo}}, \bibinfo {author} {\bibfnamefont {H.}~\bibnamefont {Firouzjahi}},
  \ and\ \bibinfo {author} {\bibfnamefont {M.}~\bibnamefont {Sasaki}},\ }\href
  {\doibase 10.1209/0295-5075/101/39001} {\bibfield  {journal} {\bibinfo
  {journal} {EPL}\ }\textbf {\bibinfo {volume} {101}},\ \bibinfo {pages}
  {39001} (\bibinfo {year} {2013})},\ \Eprint {http://arxiv.org/abs/1210.3692}
  {arXiv:1210.3692 [astro-ph.CO]} \BibitemShut {NoStop}%
\bibitem [{\citenamefont {Martin}\ \emph {et~al.}(2013)\citenamefont {Martin},
  \citenamefont {Motohashi},\ and\ \citenamefont {Suyama}}]{Martin:2012pe}%
  \BibitemOpen
  \bibfield  {author} {\bibinfo {author} {\bibfnamefont {J.}~\bibnamefont
  {Martin}}, \bibinfo {author} {\bibfnamefont {H.}~\bibnamefont {Motohashi}}, \
  and\ \bibinfo {author} {\bibfnamefont {T.}~\bibnamefont {Suyama}},\ }\href
  {\doibase 10.1103/PhysRevD.87.023514} {\bibfield  {journal} {\bibinfo
  {journal} {Phys. Rev.}\ }\textbf {\bibinfo {volume} {D87}},\ \bibinfo {pages}
  {023514} (\bibinfo {year} {2013})},\ \Eprint {http://arxiv.org/abs/1211.0083}
  {arXiv:1211.0083 [astro-ph.CO]} \BibitemShut {NoStop}%
\bibitem [{\citenamefont {Chen}\ \emph
  {et~al.}(2013{\natexlab{a}})\citenamefont {Chen}, \citenamefont {Firouzjahi},
  \citenamefont {Komatsu}, \citenamefont {Namjoo},\ and\ \citenamefont
  {Sasaki}}]{Chen:2013eea}%
  \BibitemOpen
  \bibfield  {author} {\bibinfo {author} {\bibfnamefont {X.}~\bibnamefont
  {Chen}}, \bibinfo {author} {\bibfnamefont {H.}~\bibnamefont {Firouzjahi}},
  \bibinfo {author} {\bibfnamefont {E.}~\bibnamefont {Komatsu}}, \bibinfo
  {author} {\bibfnamefont {M.~H.}\ \bibnamefont {Namjoo}}, \ and\ \bibinfo
  {author} {\bibfnamefont {M.}~\bibnamefont {Sasaki}},\ }\href {\doibase
  10.1088/1475-7516/2013/12/039} {\bibfield  {journal} {\bibinfo  {journal}
  {JCAP}\ }\textbf {\bibinfo {volume} {1312}},\ \bibinfo {pages} {039}
  (\bibinfo {year} {2013}{\natexlab{a}})},\ \Eprint
  {http://arxiv.org/abs/1308.5341} {arXiv:1308.5341 [astro-ph.CO]} \BibitemShut
  {NoStop}%
\bibitem [{\citenamefont {Pattison}\ \emph {et~al.}(2017)\citenamefont
  {Pattison}, \citenamefont {Vennin}, \citenamefont {Assadullahi},\ and\
  \citenamefont {Wands}}]{Pattison:2017mbe}%
  \BibitemOpen
  \bibfield  {author} {\bibinfo {author} {\bibfnamefont {C.}~\bibnamefont
  {Pattison}}, \bibinfo {author} {\bibfnamefont {V.}~\bibnamefont {Vennin}},
  \bibinfo {author} {\bibfnamefont {H.}~\bibnamefont {Assadullahi}}, \ and\
  \bibinfo {author} {\bibfnamefont {D.}~\bibnamefont {Wands}},\ }\href
  {\doibase 10.1088/1475-7516/2017/10/046} {\bibfield  {journal} {\bibinfo
  {journal} {JCAP}\ }\textbf {\bibinfo {volume} {1710}},\ \bibinfo {pages}
  {046} (\bibinfo {year} {2017})},\ \Eprint {http://arxiv.org/abs/1707.00537}
  {arXiv:1707.00537 [hep-th]} \BibitemShut {NoStop}%
\bibitem [{\citenamefont {Cabass}\ \emph {et~al.}(2017)\citenamefont {Cabass},
  \citenamefont {Pajer},\ and\ \citenamefont {Schmidt}}]{Cabass:2016cgp}%
  \BibitemOpen
  \bibfield  {author} {\bibinfo {author} {\bibfnamefont {G.}~\bibnamefont
  {Cabass}}, \bibinfo {author} {\bibfnamefont {E.}~\bibnamefont {Pajer}}, \
  and\ \bibinfo {author} {\bibfnamefont {F.}~\bibnamefont {Schmidt}},\ }\href
  {\doibase 10.1088/1475-7516/2017/01/003} {\bibfield  {journal} {\bibinfo
  {journal} {JCAP}\ }\textbf {\bibinfo {volume} {1701}},\ \bibinfo {pages}
  {003} (\bibinfo {year} {2017})},\ \Eprint {http://arxiv.org/abs/1612.00033}
  {arXiv:1612.00033 [hep-th]} \BibitemShut {NoStop}%
\bibitem [{\citenamefont {Aghanim}\ \emph {et~al.}()\citenamefont {Aghanim}
  \emph {et~al.}}]{Aghanim:2018eyx}%
  \BibitemOpen
  \bibfield  {author} {\bibinfo {author} {\bibfnamefont {N.}~\bibnamefont
  {Aghanim}} \emph {et~al.} (\bibinfo {collaboration} {Planck}),\ }\href@noop
  {} {\ }\Eprint {http://arxiv.org/abs/1807.06209} {arXiv:1807.06209
  [astro-ph.CO]} \BibitemShut {NoStop}%
\bibitem [{\citenamefont {Ade}\ \emph {et~al.}(2016)\citenamefont {Ade} \emph
  {et~al.}}]{Ade:2015ava}%
  \BibitemOpen
  \bibfield  {author} {\bibinfo {author} {\bibfnamefont {P.~A.~R.}\
  \bibnamefont {Ade}} \emph {et~al.} (\bibinfo {collaboration} {Planck}),\
  }\href {\doibase 10.1051/0004-6361/201525836} {\bibfield  {journal} {\bibinfo
   {journal} {Astron. Astrophys.}\ }\textbf {\bibinfo {volume} {594}},\
  \bibinfo {pages} {A17} (\bibinfo {year} {2016})},\ \Eprint
  {http://arxiv.org/abs/1502.01592} {arXiv:1502.01592 [astro-ph.CO]}
  \BibitemShut {NoStop}%
\bibitem [{\citenamefont {Pattison}\ \emph {et~al.}(2018)\citenamefont
  {Pattison}, \citenamefont {Vennin}, \citenamefont {Assadullahi},\ and\
  \citenamefont {Wands}}]{Pattison:2018bct}%
  \BibitemOpen
  \bibfield  {author} {\bibinfo {author} {\bibfnamefont {C.}~\bibnamefont
  {Pattison}}, \bibinfo {author} {\bibfnamefont {V.}~\bibnamefont {Vennin}},
  \bibinfo {author} {\bibfnamefont {H.}~\bibnamefont {Assadullahi}}, \ and\
  \bibinfo {author} {\bibfnamefont {D.}~\bibnamefont {Wands}},\ }\href@noop {}
  {\  (\bibinfo {year} {2018})},\ \Eprint {http://arxiv.org/abs/1806.09553}
  {arXiv:1806.09553 [astro-ph.CO]} \BibitemShut {NoStop}%
\bibitem [{\citenamefont {Hu}\ and\ \citenamefont {Joyce}(2017)}]{Hu:2016wfa}%
  \BibitemOpen
  \bibfield  {author} {\bibinfo {author} {\bibfnamefont {W.}~\bibnamefont
  {Hu}}\ and\ \bibinfo {author} {\bibfnamefont {A.}~\bibnamefont {Joyce}},\
  }\href {\doibase 10.1103/PhysRevD.95.043529} {\bibfield  {journal} {\bibinfo
  {journal} {Phys. Rev.}\ }\textbf {\bibinfo {volume} {D95}},\ \bibinfo {pages}
  {043529} (\bibinfo {year} {2017})},\ \Eprint
  {http://arxiv.org/abs/1612.02454} {arXiv:1612.02454 [astro-ph.CO]}
  \BibitemShut {NoStop}%
\bibitem [{\citenamefont {Cicoli}\ \emph {et~al.}(2018)\citenamefont {Cicoli},
  \citenamefont {Diaz},\ and\ \citenamefont {Pedro}}]{Cicoli:2018asa}%
  \BibitemOpen
  \bibfield  {author} {\bibinfo {author} {\bibfnamefont {M.}~\bibnamefont
  {Cicoli}}, \bibinfo {author} {\bibfnamefont {V.~A.}\ \bibnamefont {Diaz}}, \
  and\ \bibinfo {author} {\bibfnamefont {F.~G.}\ \bibnamefont {Pedro}},\ }\href
  {\doibase 10.1088/1475-7516/2018/06/034} {\bibfield  {journal} {\bibinfo
  {journal} {JCAP}\ }\textbf {\bibinfo {volume} {1806}},\ \bibinfo {pages}
  {034} (\bibinfo {year} {2018})},\ \Eprint {http://arxiv.org/abs/1803.02837}
  {arXiv:1803.02837 [hep-th]} \BibitemShut {NoStop}%
\bibitem [{\citenamefont {{\"O}zsoy}\ \emph {et~al.}(2018)\citenamefont
  {{\"O}zsoy}, \citenamefont {Parameswaran}, \citenamefont {Tasinato},\ and\
  \citenamefont {Zavala}}]{Ozsoy:2018flq}%
  \BibitemOpen
  \bibfield  {author} {\bibinfo {author} {\bibfnamefont {O.}~\bibnamefont
  {{\"O}zsoy}}, \bibinfo {author} {\bibfnamefont {S.}~\bibnamefont
  {Parameswaran}}, \bibinfo {author} {\bibfnamefont {G.}~\bibnamefont
  {Tasinato}}, \ and\ \bibinfo {author} {\bibfnamefont {I.}~\bibnamefont
  {Zavala}},\ }\href {\doibase 10.1088/1475-7516/2018/07/005} {\bibfield
  {journal} {\bibinfo  {journal} {JCAP}\ }\textbf {\bibinfo {volume} {1807}},\
  \bibinfo {pages} {005} (\bibinfo {year} {2018})},\ \Eprint
  {http://arxiv.org/abs/1803.07626} {arXiv:1803.07626 [hep-th]} \BibitemShut
  {NoStop}%
\bibitem [{\citenamefont {Byrnes}\ \emph {et~al.}()\citenamefont {Byrnes},
  \citenamefont {Cole},\ and\ \citenamefont {Patil}}]{Byrnes:2018txb}%
  \BibitemOpen
  \bibfield  {author} {\bibinfo {author} {\bibfnamefont {C.~T.}\ \bibnamefont
  {Byrnes}}, \bibinfo {author} {\bibfnamefont {P.~S.}\ \bibnamefont {Cole}}, \
  and\ \bibinfo {author} {\bibfnamefont {S.~P.}\ \bibnamefont {Patil}},\
  }\href@noop {} {\ }\Eprint {http://arxiv.org/abs/1811.11158}
  {arXiv:1811.11158 [astro-ph.CO]} \BibitemShut {NoStop}%
\bibitem [{\citenamefont {Atal}\ and\ \citenamefont
  {Germani}()}]{Atal:2018neu}%
  \BibitemOpen
  \bibfield  {author} {\bibinfo {author} {\bibfnamefont {V.}~\bibnamefont
  {Atal}}\ and\ \bibinfo {author} {\bibfnamefont {C.}~\bibnamefont {Germani}},\
  }\href@noop {} {\ }\Eprint {http://arxiv.org/abs/1811.07857}
  {arXiv:1811.07857 [astro-ph.CO]} \BibitemShut {NoStop}%
\bibitem [{\citenamefont {Saito}\ \emph {et~al.}(2008)\citenamefont {Saito},
  \citenamefont {Yokoyama},\ and\ \citenamefont {Nagata}}]{Saito:2008em}%
  \BibitemOpen
  \bibfield  {author} {\bibinfo {author} {\bibfnamefont {R.}~\bibnamefont
  {Saito}}, \bibinfo {author} {\bibfnamefont {J.}~\bibnamefont {Yokoyama}}, \
  and\ \bibinfo {author} {\bibfnamefont {R.}~\bibnamefont {Nagata}},\ }\href
  {\doibase 10.1088/1475-7516/2008/06/024} {\bibfield  {journal} {\bibinfo
  {journal} {JCAP}\ }\textbf {\bibinfo {volume} {0806}},\ \bibinfo {pages}
  {024} (\bibinfo {year} {2008})},\ \Eprint {http://arxiv.org/abs/0804.3470}
  {arXiv:0804.3470 [astro-ph]} \BibitemShut {NoStop}%
\bibitem [{\citenamefont {{Smith}}\ and\ \citenamefont
  {{LoVerde}}(2011)}]{Smith:2011}%
  \BibitemOpen
  \bibfield  {author} {\bibinfo {author} {\bibfnamefont {K.~M.}\ \bibnamefont
  {{Smith}}}\ and\ \bibinfo {author} {\bibfnamefont {M.}~\bibnamefont
  {{LoVerde}}},\ }\href {\doibase 10.1088/1475-7516/2011/11/009} {\bibfield
  {journal} {\bibinfo  {journal} {Journal of Cosmology and Astro-Particle
  Physics}\ }\textbf {\bibinfo {volume} {2011}},\ \bibinfo {eid} {009}
  (\bibinfo {year} {2011})},\ \Eprint {http://arxiv.org/abs/1010.0055}
  {arXiv:1010.0055 [astro-ph.CO]} \BibitemShut {NoStop}%
\bibitem [{\citenamefont {Byrnes}\ \emph {et~al.}(2012)\citenamefont {Byrnes},
  \citenamefont {Copeland},\ and\ \citenamefont {Green}}]{Byrnes:2012yx}%
  \BibitemOpen
  \bibfield  {author} {\bibinfo {author} {\bibfnamefont {C.~T.}\ \bibnamefont
  {Byrnes}}, \bibinfo {author} {\bibfnamefont {E.~J.}\ \bibnamefont
  {Copeland}}, \ and\ \bibinfo {author} {\bibfnamefont {A.~M.}\ \bibnamefont
  {Green}},\ }\href {\doibase 10.1103/PhysRevD.86.043512} {\bibfield  {journal}
  {\bibinfo  {journal} {Phys. Rev.}\ }\textbf {\bibinfo {volume} {D86}},\
  \bibinfo {pages} {043512} (\bibinfo {year} {2012})},\ \Eprint
  {http://arxiv.org/abs/1206.4188} {arXiv:1206.4188 [astro-ph.CO]} \BibitemShut
  {NoStop}%
\bibitem [{\citenamefont {Ferraro}\ and\ \citenamefont
  {Smith}(2015)}]{Ferraro:2014jba}%
  \BibitemOpen
  \bibfield  {author} {\bibinfo {author} {\bibfnamefont {S.}~\bibnamefont
  {Ferraro}}\ and\ \bibinfo {author} {\bibfnamefont {K.~M.}\ \bibnamefont
  {Smith}},\ }\href {\doibase 10.1103/PhysRevD.91.043506} {\bibfield  {journal}
  {\bibinfo  {journal} {Phys. Rev.}\ }\textbf {\bibinfo {volume} {D91}},\
  \bibinfo {pages} {043506} (\bibinfo {year} {2015})},\ \Eprint
  {http://arxiv.org/abs/1408.3126} {arXiv:1408.3126 [astro-ph.CO]} \BibitemShut
  {NoStop}%
\bibitem [{\citenamefont {Young}\ and\ \citenamefont
  {Byrnes}(2015)}]{Young:2015kda}%
  \BibitemOpen
  \bibfield  {author} {\bibinfo {author} {\bibfnamefont {S.}~\bibnamefont
  {Young}}\ and\ \bibinfo {author} {\bibfnamefont {C.~T.}\ \bibnamefont
  {Byrnes}},\ }\href {\doibase 10.1088/1475-7516/2015/04/034} {\bibfield
  {journal} {\bibinfo  {journal} {JCAP}\ }\textbf {\bibinfo {volume} {1504}},\
  \bibinfo {pages} {034} (\bibinfo {year} {2015})},\ \Eprint
  {http://arxiv.org/abs/1503.01505} {arXiv:1503.01505 [astro-ph.CO]}
  \BibitemShut {NoStop}%
\bibitem [{\citenamefont {Tada}\ and\ \citenamefont
  {Yokoyama}(2015)}]{Tada:2015noa}%
  \BibitemOpen
  \bibfield  {author} {\bibinfo {author} {\bibfnamefont {Y.}~\bibnamefont
  {Tada}}\ and\ \bibinfo {author} {\bibfnamefont {S.}~\bibnamefont
  {Yokoyama}},\ }\href {\doibase 10.1103/PhysRevD.91.123534} {\bibfield
  {journal} {\bibinfo  {journal} {Phys. Rev.}\ }\textbf {\bibinfo {volume}
  {D91}},\ \bibinfo {pages} {123534} (\bibinfo {year} {2015})},\ \Eprint
  {http://arxiv.org/abs/1502.01124} {arXiv:1502.01124 [astro-ph.CO]}
  \BibitemShut {NoStop}%
\bibitem [{\citenamefont {Adshead}\ \emph {et~al.}(2013)\citenamefont
  {Adshead}, \citenamefont {Hu},\ and\ \citenamefont
  {Miranda}}]{Adshead:2013zfa}%
  \BibitemOpen
  \bibfield  {author} {\bibinfo {author} {\bibfnamefont {P.}~\bibnamefont
  {Adshead}}, \bibinfo {author} {\bibfnamefont {W.}~\bibnamefont {Hu}}, \ and\
  \bibinfo {author} {\bibfnamefont {V.}~\bibnamefont {Miranda}},\ }\href
  {\doibase 10.1103/PhysRevD.88.023507} {\bibfield  {journal} {\bibinfo
  {journal} {Phys. Rev.}\ }\textbf {\bibinfo {volume} {D88}},\ \bibinfo {pages}
  {023507} (\bibinfo {year} {2013})},\ \Eprint {http://arxiv.org/abs/1303.7004}
  {arXiv:1303.7004 [astro-ph.CO]} \BibitemShut {NoStop}%
\bibitem [{\citenamefont {Weinberg}(2005)}]{Weinberg:2005vy}%
  \BibitemOpen
  \bibfield  {author} {\bibinfo {author} {\bibfnamefont {S.}~\bibnamefont
  {Weinberg}},\ }\href {\doibase 10.1103/PhysRevD.72.043514} {\bibfield
  {journal} {\bibinfo  {journal} {Phys. Rev.}\ }\textbf {\bibinfo {volume}
  {D72}},\ \bibinfo {pages} {043514} (\bibinfo {year} {2005})},\ \Eprint
  {http://arxiv.org/abs/hep-th/0506236} {arXiv:hep-th/0506236 [hep-th]}
  \BibitemShut {NoStop}%
\bibitem [{\citenamefont {Adshead}\ \emph
  {et~al.}(2009{\natexlab{a}})\citenamefont {Adshead}, \citenamefont
  {Easther},\ and\ \citenamefont {Lim}}]{Adshead:2009cb}%
  \BibitemOpen
  \bibfield  {author} {\bibinfo {author} {\bibfnamefont {P.}~\bibnamefont
  {Adshead}}, \bibinfo {author} {\bibfnamefont {R.}~\bibnamefont {Easther}}, \
  and\ \bibinfo {author} {\bibfnamefont {E.~A.}\ \bibnamefont {Lim}},\ }\href
  {\doibase 10.1103/PhysRevD.80.083521} {\bibfield  {journal} {\bibinfo
  {journal} {Phys. Rev.}\ }\textbf {\bibinfo {volume} {D80}},\ \bibinfo {pages}
  {083521} (\bibinfo {year} {2009}{\natexlab{a}})},\ \Eprint
  {http://arxiv.org/abs/0904.4207} {arXiv:0904.4207 [hep-th]} \BibitemShut
  {NoStop}%
\bibitem [{\citenamefont {Adshead}\ \emph
  {et~al.}(2009{\natexlab{b}})\citenamefont {Adshead}, \citenamefont
  {Easther},\ and\ \citenamefont {Lim}}]{Adshead:2008gk}%
  \BibitemOpen
  \bibfield  {author} {\bibinfo {author} {\bibfnamefont {P.}~\bibnamefont
  {Adshead}}, \bibinfo {author} {\bibfnamefont {R.}~\bibnamefont {Easther}}, \
  and\ \bibinfo {author} {\bibfnamefont {E.~A.}\ \bibnamefont {Lim}},\ }\href
  {\doibase 10.1103/PhysRevD.79.063504} {\bibfield  {journal} {\bibinfo
  {journal} {Phys. Rev.}\ }\textbf {\bibinfo {volume} {D79}},\ \bibinfo {pages}
  {063504} (\bibinfo {year} {2009}{\natexlab{b}})},\ \Eprint
  {http://arxiv.org/abs/0809.4008} {arXiv:0809.4008 [hep-th]} \BibitemShut
  {NoStop}%
\bibitem [{\citenamefont {Passaglia}\ and\ \citenamefont
  {Hu}(2018)}]{Passaglia:2018afq}%
  \BibitemOpen
  \bibfield  {author} {\bibinfo {author} {\bibfnamefont {S.}~\bibnamefont
  {Passaglia}}\ and\ \bibinfo {author} {\bibfnamefont {W.}~\bibnamefont {Hu}},\
  }\href {\doibase 10.1103/PhysRevD.98.023526} {\bibfield  {journal} {\bibinfo
  {journal} {Phys. Rev.}\ }\textbf {\bibinfo {volume} {D98}},\ \bibinfo {pages}
  {023526} (\bibinfo {year} {2018})},\ \Eprint
  {http://arxiv.org/abs/1804.07741} {arXiv:1804.07741 [astro-ph.CO]}
  \BibitemShut {NoStop}%
\bibitem [{\citenamefont {Chen}\ \emph
  {et~al.}(2013{\natexlab{b}})\citenamefont {Chen}, \citenamefont {Firouzjahi},
  \citenamefont {Namjoo},\ and\ \citenamefont {Sasaki}}]{Chen:2013aj}%
  \BibitemOpen
  \bibfield  {author} {\bibinfo {author} {\bibfnamefont {X.}~\bibnamefont
  {Chen}}, \bibinfo {author} {\bibfnamefont {H.}~\bibnamefont {Firouzjahi}},
  \bibinfo {author} {\bibfnamefont {M.~H.}\ \bibnamefont {Namjoo}}, \ and\
  \bibinfo {author} {\bibfnamefont {M.}~\bibnamefont {Sasaki}},\ }\href
  {\doibase 10.1209/0295-5075/102/59001} {\bibfield  {journal} {\bibinfo
  {journal} {EPL}\ }\textbf {\bibinfo {volume} {102}},\ \bibinfo {pages}
  {59001} (\bibinfo {year} {2013}{\natexlab{b}})},\ \Eprint
  {http://arxiv.org/abs/1301.5699} {arXiv:1301.5699 [hep-th]} \BibitemShut
  {NoStop}%
\end{thebibliography}%

\end{document}